\documentclass[12pt,aps,showpacs,nofootinbib,showkeys]{revtex4-2}

\usepackage{amsmath,amssymb,amsfonts}
\usepackage{graphicx,graphics}
\usepackage{bm}
\usepackage{color}
\usepackage{epstopdf}
\usepackage{setspace}
\usepackage{braket}
\usepackage{verbatim}
\usepackage{cases}
\usepackage{empheq}
\usepackage{pgfplots}
\usepackage{hyperref}
\usepackage{multirow}

\begin{document}

\title{Low-Mass Neutron Stars and Effective Phase Transitions from a Hybrid Van der Waals–Polytropic Equation of State}

\author{P. H. F. Arruda}
\author{S. D. Campos}\email{sergiodc@ufscar.br}
\address{Applied Mathematics Laboratory -- CCTS/DFQM, Federal University of S\~ao Carlos, Sorocaba, CEP 18052-780, Brazil}

\begin{abstract}
We study phase–transition–like behavior in neutron stars using a simplified, piecewise equation of state that couples a modified van der Waals–type core to a polytropic crust. The model remains analytically tractable while allowing for nonlinear density dependence. We impose thermodynamic and causal consistency conditions and determine the critical densities at which the curvature of the pressure–energy-density relation changes. In the non-relativistic limit, the generalized Lane–Emden equations describe a smooth core–crust transition layer. We integrate the Tolman–Oppenheimer–Volkoff equations across different $(\tau_1,\sigma_1)$ regimes, where these parameters encode thermal and interaction effects in the core. The resulting mass–radius sequences yield low neutron star masses $(0.99-2.05)M_{\odot}$, and the chemical potential exhibits the characteristic signatures of phase-transition behavior at densities well above the matching point. Our results show that analytic EOS models can reproduce the key phenomenology of phase transitions and provide a controlled framework for exploring low-mass neutron star configurations.
\end{abstract}

\pacs{97.60.Jd; 26.60.+c; 26.60.-c}
\keywords{neutron star; equation of state; phase transition}

\maketitle

\section{Introduction}
\label{sec:intro}

Since the derivation of the Tolman–Oppenheimer–Volkoff (TOV) equation \cite{R.C.Tolman.1934,R.C.Tolman.1939,R.J.Oppenheimer.G.M.Volkoff.1939}, our understanding of the structure and internal physics of neutron stars has improved significantly. Early attempts to describe ultra-dense matter were essentially non-relativistic, dating back to Fermi’s work on the degenerate quantum gas \cite{E.Fermi.1926}. The TOV equation introduced the first fully relativistic treatment of hydrostatic equilibrium for compact stars, along with an early relativistic upper bound for the neutron star (NS) mass of $\sim 0.7\,M_\odot$ in the original Oppenheimer–Volkoff model.

Observations have since revealed NS masses far exceeding this bound, with several precisely measured heavy pulsars, such as PSR J0952–0607 $(2.35\pm0.17)\,M_\odot$ \cite{C.G.Bassa.et.al.2017}, PSR J0348+0432 $(2.01\pm0.04)\,M_\odot$ \cite{J.Antoniadis.et.al.2013}, and the massive system discussed in Ref.~\cite{M.Linares.Astrophys.J.2018}. The mass/radius relation is crucial for determining the correct equation of state (EoS), as can be viewed in the recent analysis of NICER data: PSR J0030+0451 \cite{Miller.2019,Riley.2019} and PSR J0740+6620 \cite{Miller.2021,Riley.2021}.

On the other end of the mass spectrum, the question of the \emph{minimum} NS mass remains open. Recent observations have reported NS candidates with masses as low as $\sim 0.77\,M_\odot$ \cite{V.Doroshenko.NatureAstronomy.1-9.2022,J.E.Horvath.AA.672.L11.2023} and $\sim (0.98\pm0.03)\,M_\odot$ \cite{J.Lin.2023}. Furthermore, three-dimensional simulations of core-collapse supernovae suggest that NS remnants can form with gravitational masses below $\sim 1.192\,M_\odot$ \cite{B.Muller.Phys.Rev.Lett.134.071403.2025,Wang.2024} (and references therein). This value is intriguingly close to the Chandrasekhar limit for white dwarfs, though the underlying physical mechanisms are entirely different. The GW170817 mass constraints may shed light on the minimum-mass mechanism \cite{Abbott.2018,Abbott.2019}.

One of the main challenges in high-energy and nuclear astrophysics is to characterize dense nuclear matter under conditions that are inaccessible to terrestrial experiments. The pressure and energy density in the interior of NSs far exceed those attained in current laboratories, and the only input required to solve the TOV equations is the EoS of cold, $\beta$-equilibrated matter. Consequently, the construction of realistic EoS at supra-nuclear densities is a central and still controversial topic.

Since the 1950s, many EoS models have been developed, employing a variety of many-body techniques and nuclear interaction schemes. Cameron \cite{A.G.W.Cameron.1959} provided one of the first descriptions of NS matter as a dense nuclear system. Baym, Pethick, and Pines \cite{G.Baym.1969} investigated neutron superfluidity and its impact on cooling and magnetic-field evolution, while Baym, Pethick, and Sutherland \cite{G.Baym.1971} introduced an influential EoS that includes nuclear interactions at subnuclear densities. A widely used microphysical EoS is that of Akmal, Pandharipande, and Ravenhall \cite{A.Akmal.1998}. Reviews such as \cite{A.R.Raduta.2021,Z.Ji.2025} emphasize the diversity of available EoS and their respective strengths and limitations.

Despite this variety, relatively few EoS are sufficiently simple (analytical or semi-analytical) to allow a systematic exploration of phase-transition-like behavior while remaining tractable in both low- and high-density regimes. This motivates the construction of simplified, piecewise EoS that can encode the qualitative features of phase transitions and provide insight into the internal structure of NSs. In particular, such models can help address questions about the existence of low-mass configurations and the location (and nature) of possible phase transitions in the core.

The use of van der Waals (vdW)–type EoS in hadronic and quark matter contexts is long-standing \cite{J.Baacke.1977}. In many approaches, a vdW EoS is combined with MIT bag-like models \cite{A.Chodos.1974} or with Hagedorn-type descriptions of hadron resonance gases, where non-interacting hadrons form a multi-component gas dominated by resonance formation (see \cite{Typel.2016,V.Vovchenko.2020} and references therein). More broadly, phase transitions in EoS are of fundamental interest because they encode abrupt changes in the microscopic state of matter that may leave observable imprints in macroscopic systems, from laboratory fluids to compact stars \cite{verma.arxiv.2025}. 

In this work, we take a complementary approach and investigate whether a minimally parameterized, analytically tractable equation of state can already
encode key signatures normally attributed to phase transitions in dense matter. To this end, we construct a piecewise model in which the stellar core is described by a modified van der Waals–type expression, and the outer layer is described by a simple polytropic form. The resulting EoS is not intended to reproduce detailed microphysics; rather, it provides a controlled setting in which the curvature of $p(\varepsilon)$, the existence of critical densities, and the structure of the core–crust interface can be analyzed systematically.

A central feature of our formulation is that the two core parameters $(\tau_{1},\sigma_{1})$ control, respectively, the density dependence of an
effective repulsive (“thermal”) contribution and of an interaction-driven high-density term. Mapping the model in this parameter space reveals distinct
regimes characterized by the number and location of critical densities, which we interpret as phase-transition-like domains. We then quantify how these
regimes affect the non-relativistic limit, the structure of the transition layer, the relativistic stellar solutions, and the behavior of the baryonic
chemical potential.

Despite the many available microphysical equations of state, relatively few analytic models allow for a controlled and transparent exploration of phase–transition–like behavior within the TOV framework. Fully realistic EoS frequently involve large numbers of parameters, necessitating numerical interpolation or lacking any closed-form analytic representation. These features hinder the identification of the specific microphysical components that drive softening, stiffening, or the emergence of critical phenomena. In contrast, analytically tractable models facilitate a more transparent examination of how specific nonlinear density dependence governs global stellar structure and determine the onset and location of critical densities within the stellar core. The purpose of this work is therefore not to reproduce a realistic nuclear model, but to provide a simplified, fully analytic testbed that reveals how a small number of parameters can control the emergence of phase-transition signatures and low-mass neutron-star configurations. 

The paper is organized as follows. In Section~\ref{sec:model}, we introduce the piecewise EoS, along with consistency constraints such as thermal stability and causality, and we derive critical densities. Section~\ref{sec:nonrel} is devoted to the non-relativistic treatment, where we derive generalized Lane–Emden–type equations and analyze the core–crust interface. Section~\ref{sec:rel} presents the relativistic treatment through the TOV equations, together with numerical results for various parameter regimes, including mass–radius curves and chemical potentials. We summarize and discuss our results, including observational implications and possible extensions, in Section~\ref{sec:final}.

\section{Equation of State}
\label{sec:model}

We consider a non-rotating, electrically neutral, spherically symmetric neutron star in hydrostatic equilibrium. To be physically acceptable, the model must satisfy several standard conditions:

\begin{itemize}
    \item[1.] \emph{Thermal stability}: the pressure must increase with energy density,
    \begin{equation}
        \frac{dp}{d\varepsilon} \ge 0;
    \end{equation}
    \item[2.] \emph{Causality}: the adiabatic sound speed must not exceed the speed of light, 
    \begin{equation}
        v_s^2 \equiv \frac{dp}{d\varepsilon} \le c^2;
    \end{equation}
    \item[3.] \emph{Regularity}: the pressure should vanish when the density vanishes,
    \begin{equation}
        p(\varepsilon=0) = 0.
    \end{equation}
\end{itemize}

In principle, an NS model starts from a given EoS that encapsulates the microphysics of neutrons, protons, electrons, and possibly more exotic constituents. Analytical solutions to the TOV equations are rare due to their nonlinear structure. Tolman \cite{R.C.Tolman.1939} derived eight exact solutions; however, only one yields a physically reasonable NS configuration and reproduces the classical Oppenheimer–Volkoff mass limit $\sim 0.7\,M_\odot$, which was later superseded by more realistic $\sim 2\,M_\odot$ limits \cite{A.G.W.Cameron.1959}.

Quantum effects play a central role in NS matter. At the densities of interest, the de Broglie wavelength $\lambda = h/(\gamma m v)$ can exceed the inter-particle spacing, implying strong degeneracy and possibly non-trivial phases of Quantum Chromodynamics. Typical NS temperatures $T\sim10^9\,\text{K}$ are much lower than the Fermi temperature $T_F \sim 10^{12}$-$10^{13}\,\text{K}$, so neutrons form a cold, degenerate Fermi gas. Thus, the star can be approximated as a cold, compact object.

\begin{figure}[t]
    \centering
    \includegraphics[width=0.4\linewidth]{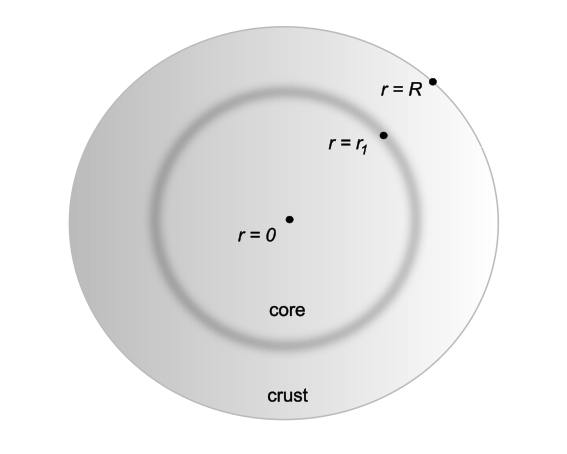}
    \caption{Schematic, not-to-scale representation of the NS interior as modeled here. The stellar core ($\varepsilon > \varepsilon_0$) is characterized by a modified vdW–type EoS, whereas the crustal region ($\varepsilon \leq \varepsilon_0$) is represented by a two-term polytropic EoS. The transition region around $r_1$ is the transition layer, where the two descriptions are approximately valid.}
    \label{fig:layers_ns}
\end{figure}

For simplicity, we assume the constituents can be described as an electrically neutral, relativistic, free Fermi gas in chemical equilibrium, and we treat the NS as having a characteristic radius $R$, which allows us to partition the interior into two regions according to the energy density:
\begin{equation}
    \text{crust: } \varepsilon \le \varepsilon_0, \qquad
    \text{core: } \varepsilon > \varepsilon_0,
\end{equation}
where $\varepsilon_0$ is a reference (saturation) density. The conceptual structure is shown in Fig.~\ref{fig:layers_ns}. The blurred region around $r_1$ represents an interface where phase-transition behavior may occur.

We adopt natural units $c=G=k_B=\hbar=1$ throughout. In the core, we introduce a modified vdW-type EoS of the form
\begin{equation}
    p = p(\varepsilon)
    = \frac{a\,\varepsilon_0\,T}{(\varepsilon_0/\varepsilon)-1}
      + \beta_1\left(\frac{\varepsilon}{\varepsilon_0}\right)^{\sigma_1},
    \qquad \varepsilon_0 < \varepsilon,
    \label{eq:press1}
\end{equation}
where $\varepsilon(r)$ is the energy density, $\beta_1$ is a parameter describing the strength of the (non-thermal) interaction, and $\sigma_1$ is a dimensionless exponent. We identify $\varepsilon_0$ with the saturation density, and $a$ is a real parameter with units of pressure divided by (temperature $\times$ density). Without loss of generality, we set $\varepsilon_0 = 1$ in the following.

In standard NS modeling, the interior is often considered to be approximately isothermal, with an effective redshifted temperature $T^\infty$ determined by the star's cooling history. Here, we relax this assumption and allow for a simple density-dependent temperature,
\begin{equation}
    T(\varepsilon) = T_0\left(\frac{\varepsilon}{\varepsilon_0}\right)^{\tau_1},
    \label{eq:temp1}
\end{equation}
where $\tau_1$ is a dimensionless parameter, and $T_0$ is a reference temperature. We typically consider $|\tau_1|\ll 1$, such that $T$ deviates only mildly from $T_0$:
\begin{equation}
    \frac{(\varepsilon/\varepsilon_0)^{\tau_1}-1}{(\varepsilon/\varepsilon_0)^{\tau_1}} \approx 0.
\end{equation}

Combining Eqs.~\eqref{eq:press1} and \eqref{eq:temp1}, we obtain the core EoS
\begin{equation}
    p(\varepsilon)
    = \frac{\alpha_1}{(\varepsilon_0/\varepsilon)-1}
      \left(\frac{\varepsilon}{\varepsilon_0}\right)^{\tau_1}
      + \beta_1\left(\frac{\varepsilon}{\varepsilon_0}\right)^{\sigma_1},
    \qquad \varepsilon_0<\varepsilon,
    \label{eq:press2}
\end{equation}
where $\alpha_1 = a\varepsilon_0 T_0$.

In the crust, where $\varepsilon/\varepsilon_0 \le 1$, we adopt a two-term polytropic EoS,
\begin{equation}
    p(\varepsilon) = \alpha_2\left(\frac{\varepsilon}{\varepsilon_0}\right)^{\tau_2}
                  + \beta_2\left(\frac{\varepsilon}{\varepsilon_0}\right)^{\sigma_2},
    \qquad \varepsilon \le \varepsilon_0,
\end{equation}
with $\tau_2\neq \sigma_2$ and $\alpha_2$ absorbing any effective temperature dependence in the crust. The piecewise EoS can be compactly written as
\begin{empheq}[left={p(\varepsilon) = \empheqlbrace}]{align}
\label{eq:sist1}
  \dfrac{\alpha_1}{(\varepsilon_0/\varepsilon-1)}
  \left(\dfrac{\varepsilon}{\varepsilon_0}\right)^{\tau_1}
  + \beta_1\left(\dfrac{\varepsilon}{\varepsilon_0}\right)^{\sigma_1},
  &\qquad \varepsilon_0 < \varepsilon,\\[0.3em]
\label{eq:sist2}
  \alpha_2\left(\dfrac{\varepsilon}{\varepsilon_0}\right)^{\tau_2}
  + \beta_2\left(\dfrac{\varepsilon}{\varepsilon_0}\right)^{\sigma_2},
  &\qquad \varepsilon \le \varepsilon_0.
\end{empheq}

The first term in each branch represents the (effective) thermal contribution, whereas the second term encodes non-thermal interactions. At the center $r=0$, we impose $\varepsilon(0)=\varepsilon_f$ (finite central density), with the corresponding $p_f=p(\varepsilon_f)$. At the surface $r=R$, we require $\varepsilon(R)=0$ and $p(R)=0$. 

The parameter $\alpha_1$ represents the repulsive term, which controls how strongly matter resists compression at densities above $\varepsilon_0$. A large positive $\alpha_1$ indicates a strong repulsion regime, while a large negative $\alpha_1$ reduces it, representing a softening NS and allowing for a phase transition. The parameter $\beta_1$ represents the strength of high-density interaction, which is a measure of the multi-body interaction effects at energy densities far from $\varepsilon_0$. For large $\beta_1$, the pressure increases rapidly, supporting larger NS masses. However, if it is arbitrarily large, it can violate causality.

The exponent $\tau_1$ controls the rate at which thermal repulsion grows ($\tau_1>0$, stiffens the EoS) or decreases ($\tau_1<0$, softens the EoS). On the other hand, the exponent $\sigma_1$ represents the effective compressibility of matter in the ultra-dense regime: $\sigma_1>0$ contributes to very stiff matter that supports a heavy NS ($>2M_{\odot}$). When $\sigma\leq 0$, one has a soft EoS with possible mixed phase transitions.

The continuity of pressure and energy density across the core–crust transition is not enforced exactly since the interface is modeled as a blurred region around a radius $r_1$. Instead, we require approximate matching across a small interval containing $r_1$: 
\begin{equation}
\begin{aligned}
   &\left[
      \frac{\alpha_1}{(\varepsilon_0/\varepsilon_+ - 1)}
      \left(\frac{\varepsilon_+}{\varepsilon_0}\right)^{\tau_1}
      + \beta_1\left(\frac{\varepsilon_+}{\varepsilon_0}\right)^{\sigma_1}
   \right]
   - \left[
      \alpha_2\left(\frac{\varepsilon_-}{\varepsilon_0}\right)^{\tau_2}
      + \beta_2\left(\frac{\varepsilon_-}{\varepsilon_0}\right)^{\sigma_2}
   \right] \approx \delta \ll 1,
\end{aligned}
\label{eq:border}
\end{equation}
where $\varepsilon_0/\varepsilon_+\ll 1$, $\varepsilon_0<\varepsilon_+$ (core side), and $\varepsilon_-\le \varepsilon_0$ (crust side). The small quantity $\delta$ measures the residual pressure mismatch across the interface.

\subsection{Thermal stability and critical densities}

Thermal stability requires $dp/d\varepsilon\ge 0$, which in our model yields an inequality involving the ratio $\beta_{1}\sigma_{1}/\alpha_{1}$ and the relative difference $(\varepsilon/\varepsilon_{0}-1)$. For the core branch \eqref{eq:sist1}, a direct calculation followed by the substitution $\varepsilon = (1+x)\varepsilon_0$ yields the inequality
\begin{equation}\label{eq:inequa13}
    (1+x)^{\tau_1-\sigma_1+2}(\tau_1 x-1)\leq \frac{\beta_1\sigma_1}{\alpha_1\varepsilon_2^2}=\frac{\beta_1\sigma_1}{\alpha_1}
\end{equation}
with $\sigma_1\ne\tau_1$ to avoid divergences. Inequality \eqref{eq:inequa13} shows that the existence of real-valued, thermally stable solutions depends sensitively on the relative magnitudes and signs of the parameters $(\alpha_1,\beta_1,\sigma_1,\tau_1)$. In particular, $\sigma_1\neq\tau_1$ implies that thermal and non-thermal contributions cannot scale identically with $\varepsilon$, which is crucial if one wants to allow for changes in the stiffness (or softness) of the EoS and for the possibility of phase transitions or metastable branches. Notice, moreover, that the inequality \eqref{eq:inequa13} controls where the pressure ceases to be monotonic, indicating where the concavity changes sign, which is typical of phase transitions.

The core EoS admits a critical energy density $\varepsilon_c$ where the slope and curvature of $p(\varepsilon)$ satisfy specific conditions. One finds the critical energy density
\begin{equation}
\frac{\varepsilon_c}{\varepsilon_0}
= \frac{\sigma_1(2\tau_1+1) - 2\tau_1(\tau_1+1) - 1
  \pm \sqrt{(\sigma_1+1)^2 - 4\tau_1(\tau_1-\sigma_1+1)}}
  {2\tau_1(\sigma_1-\tau_1)},
\label{eq:critico}
\end{equation}
which is real and positive only if
\begin{equation}
  0 \leq (\sigma_1+1)^2 - 4\tau_1(\tau_1-\sigma_1+1),
  \label{eq:ineq2s}
\end{equation}
with $\sigma_1\neq\tau_1$. From Eq.~\eqref{eq:ineq2s}, one can derive the condition
\begin{equation}
    \sigma_1 \geq -1 - 2\tau_1 + 2\sqrt{2\tau_1(\tau_1+1)}.
\end{equation}

\begin{figure}
    \centering
    \includegraphics[width=0.5\linewidth]{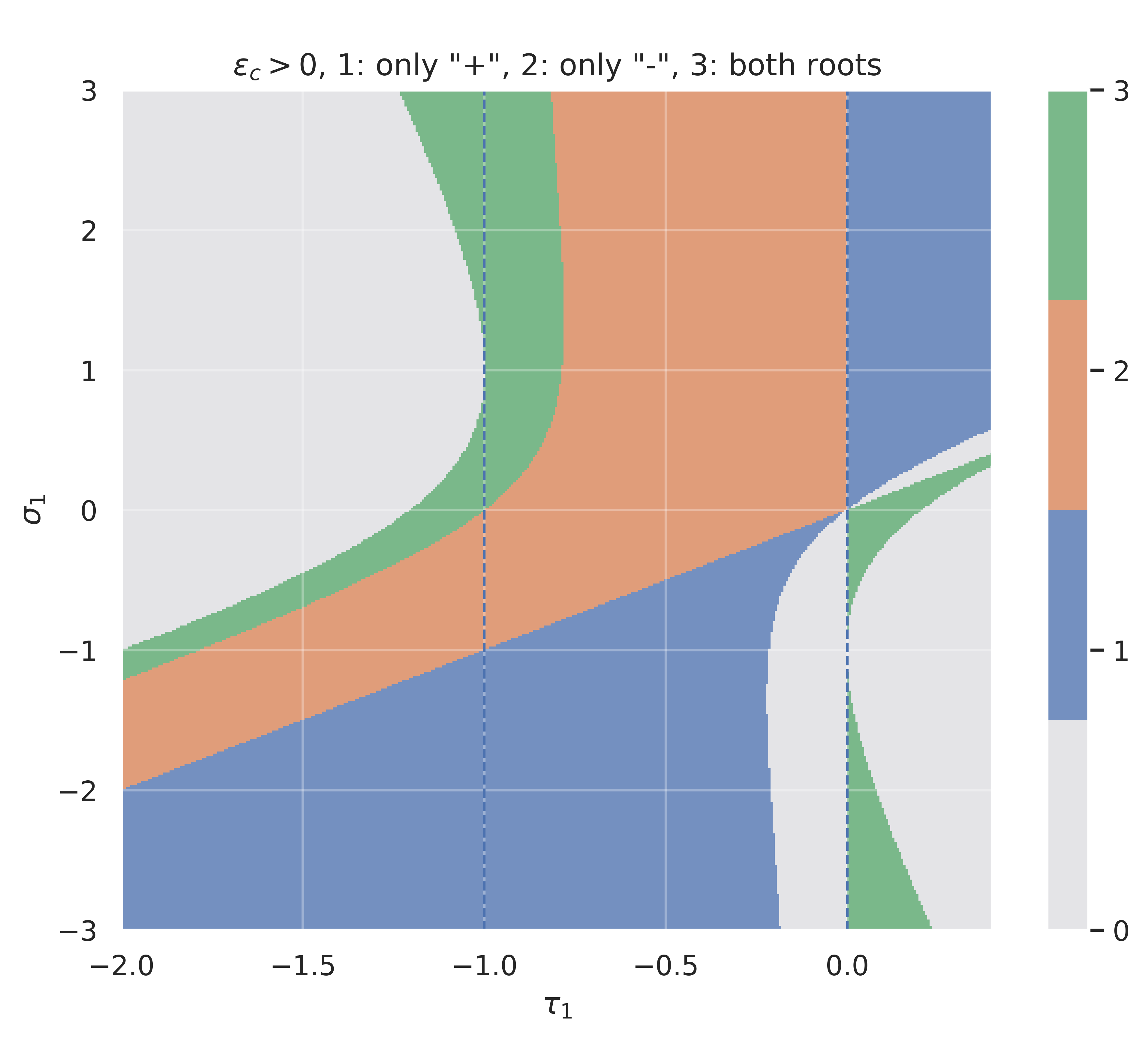}
    \caption{Plot of the parameter space ($\tau_1,\sigma_1$) showing the validity region where $\varepsilon_c>0$.}
    \label{fig:mapa_parametros}
\end{figure}

The above result \eqref{eq:critico} marks changes in the curvature of $p(\varepsilon)$ and indicates the onset of phase-transition-like behavior. The sign of the discriminant determines the number of real roots, while the signs of the numerator and denominator determine which roots have positive densities. The resulting classification is summarized in Fig.~\ref{fig:mapa_parametros}: the EoS may admit zero, one, or two positive critical densities, leading to monotonic, weakly nonlinear, or strongly spinodal-like behavior, respectively.

For $\tau_1\leq -1$ or $0<\tau_1$, there exist values of $\sigma_1$ such that the $+$ sign in Eq.~\eqref{eq:critico} yields a positive $\varepsilon_c$. However, when both exponents are negative, the resulting configurations fail to produce a physically viable NS (e.g., due to pathological behavior at high densities), so the case $\tau_1\leq -1$ with $\sigma_1<0$ is retained only as a mathematical possibility.

For the minus sign in Eq.~\eqref{eq:critico}, a positive $\varepsilon_c$ is obtained in two intervals of $\tau_1$: for $0<\tau_1\lesssim 0.142$, one finds $\sigma_1<0$; while for $0.143\lesssim\tau_1\lesssim 0.207$, one has $\sigma_1>0$. These regimes encode different qualitative balances between thermal and interaction pressures and are summarized in Table~\ref{tab:regimes}, where the unphysical case $\tau_1\leq -1$ with $\sigma_1<0$ is explicitly noted.

\begin{table}[t]
    \centering
    \begin{tabular}{|c|c|}
    \hline
    \multicolumn{2}{|c|}{Core regimes in the $(\tau_1,\sigma_1)$ plane} \\
    \hline 
    $\tau_1\leq -1$    & $0\leq \sigma_1$ \\
    \hline
    $0< \tau_1$     & $\exists\,\sigma_1$ s.t. $\varepsilon_c>0$ \\
    \hline
    $0<\tau_1\lesssim 0.142$ & $\sigma_1<0$ \\
    \hline
    $0.143\lesssim \tau_1\lesssim 0.207$ & $\sigma_1>0$   \\
    \hline
    \end{tabular}
    \caption{Qualitative regimes for the $(\tau_1,\sigma_1)$ parameter space based on the positivity and reality of the critical density $\varepsilon_c$ obtained from Eq.~\eqref{eq:critico}. The line $\tau_1\leq -1,\sigma_1<0$ is mathematically allowed but leads to unphysical NS configurations.}
    \label{tab:regimes}
\end{table}

Figure~\ref{fig:mapa_parametros} shows the classification of the parameter space according to the number of positive critical densities obtained from Eq.~\eqref{eq:critico}. Regions with no positive roots correspond to monotonic and comparatively stiff
equations of state. When a single positive root exists, the EoS develops a mild change in curvature, typically associated with softening at intermediate
densities. The region with two positive roots is the most nonlinear; the pressure exhibits two inflection points. It resembles a spinodal-like structure,
despite the absence of a true phase transition in the model. These domains, therefore, represent effective ``phase-transition-like'' regimes. This classification provides a direct link between the analytic parameters 
$(\tau_{1},\sigma_{1})$ and the qualitative physical behavior of the EoS.

The regime with no positive, critical root corresponds to the region where no physically meaningful critical energy density exists: 
\begin{eqnarray}\nonumber
    \Delta=(\sigma_1+1)^2-4\tau_1(\tau_1-\sigma_1+1)
\end{eqnarray}
is negative or imaginary. The EoS is strictly monotonic and does not exhibit any indications of non-ideal or interaction-induced effects of sufficient magnitude to produce thermodynamic signatures characteristic of a phase transition. In NS matter, it provides a simple and effective EoS characterized by smoothly varying compressibility, and it does not exhibit any internal first-order phase transitions. 

In the plus sign case, we have the upper branch of the critical energy density, corresponding to the stiffening side of the EoS. The pressure profile $p(\varepsilon)$ exhibits an inflection-like feature of physical relevance; however, its magnitude is insufficient to yield a fully developed S-shaped curve. In this regime, the EoS incorporates weakly non-ideal intermolecular interactions and the onset of many-body effects. Within an NS, this regime may correspond to an EoS with soft-to-moderately stiff characteristics, and we do not expect strong first-order phase transitions. Instead, it is potentially consistent with a single, continuous crossover from nuclear to high-density matter. 

The minus sign indicates the lower-branch solution. This root typically occurs near the onset of pronounced attractive interactions in the modified vdW term; that is, in the regime where the negative contribution to the equation of state is dominant. The EoS is more strongly influenced by the interaction term $\beta_1(\varepsilon/\varepsilon_0)^{\sigma_1}$, suggesting softening at intermediate densities. This regime may signify the emergence of a subdominant metastable phase or the development of a localized concave curvature in the underlying problem. Within the framework of NS physics, this region may correspond to a class of models that exhibit a ``hidden'' regime of reduced stiffness in the EoS, for example, as a precursor to quark-matter–induced softening, without a genuine phase transition. 

The final scenario, characterized by two positive critical roots, is of particular physical interest. It is mathematically analogous to a vdW thermodynamic system in which the presence of two stationary points delineates a spinodal region. This region signals mechanical instability and suggests the possibility of phase coexistence. For this case, the EoS develops non-monotonic regions in the $dp/d\varepsilon$ plot, exhibiting potential S-shaped $p(\varepsilon)$ behavior. Moreover, it constitutes an unambiguous signature of a first-order phase transition, characterized by the occurrence of a mixed-phase regime or, at a minimum, a metastable state with intermediate density. This regime can be associated with a variety of phase transitions, including those from purely nucleonic to hyperonic matter, from nucleonic to deconfined quark matter, hadron–quark mixed phases described within the Gibbs construction, and pasta-like or otherwise clustered states at intermediate densities. For pedagogical purposes, we construct Table \ref{tab:pedagogico}, which delineates the roles of $\tau_1$ and $\sigma_1$. 
\begin{table}[]
    \centering
    \begin{tabular}{|c|c|c|}
\hline
Parameter & Controls & Physical Meaning \\
   \hline
$\alpha_1$ & amplitude (repulsive term) & short-range repulsion/excluded-volume strength \\
$\beta_1$  &amplitude (interaction term) & high-density multi-body effects     \\
$\tau_1$   & density (repulsion) & intermediate-density stiffness or softening    \\
$\sigma_1$ & density (interactions) & ultra-high-density stiffness of softening \\
\hline
    \end{tabular}
    \caption{Physical role of each parameter in vdW EoS.}
    \label{tab:pedagogico}
\end{table}

For the crust branch \eqref{eq:sist2}, the condition $dp/d\varepsilon=0$ leads to a critical density
\begin{equation}
    \frac{\varepsilon_c}{\varepsilon_0}
    = \left(\frac{\beta_2}{\alpha_2}\right)^{\tau_2-\sigma_2},
    \label{eq:critico2}
\end{equation}
provided $\beta_2/\alpha_2 >0$. For simplicity, we assume $\alpha_2$ and $\beta_2$ share the same sign. If $\tau_2=\sigma_2$ and $\beta_2\neq 0$, the condition $\varepsilon_c=\varepsilon_0$ holds, implying that thermal and non-thermal contributions in the crust scale identically with density and cannot be distinguished in their leading dependence. If $\beta_2>\alpha_2$, then $\tau_2>\sigma_2$ is excluded because it would force $\varepsilon_c>\varepsilon_0$, contradicting $\varepsilon\leq\varepsilon_0$ in the crust. Conversely, if $\alpha_2>\beta_2$, then $\sigma_2>\tau_2$ is forbidden. In summary, for $\beta_2>\alpha_2$, one must have $\sigma_2>\tau_2$, while for $\alpha_2>\beta_2$, one finds $\tau_2>\sigma_2$.

The critical densities \eqref{eq:critico} and \eqref{eq:critico2} characterize points where the competition between different pressure contributions changes qualitatively. They define structural reference points within the internal profile at which the EoS transitions into a distinct effective regime. For later use, it is convenient to introduce dimensionless pressures by normalizing to the respective critical pressures $p_1$ and $p_2$:
\begin{empheq}[left={\hat{p}(\varepsilon) = \empheqlbrace}]{align}
\label{eq:sistdimensionless1}  
  \frac{1}{p_1}\left[
    \frac{\alpha_1}{(\varepsilon_0/\varepsilon-1)}
    \left(\frac{\varepsilon}{\varepsilon_0}\right)^{\tau_1}
    + \beta_1\left(\frac{\varepsilon}{\varepsilon_0}\right)^{\sigma_1}
  \right],
  &\qquad \varepsilon_0 < \varepsilon,\\[0.3em]
\label{eq:sistdimensionless2} 
  \frac{1}{p_2}\left[
    \alpha_2\left(\frac{\varepsilon}{\varepsilon_0}\right)^{\tau_2}
    + \beta_2\left(\frac{\varepsilon}{\varepsilon_0}\right)^{\sigma_2}
  \right],
  &\qquad \varepsilon\leq \varepsilon_0.
\end{empheq}

\section{Non-Relativistic Treatment}
\label{sec:nonrel}

Before considering the fully relativistic TOV equations, it is useful to examine the non-relativistic limit, which leads to generalized Lane–Emden–type equations. This provides analytical insight into how the piecewise EoS shapes the internal structure and the effective behavior of phase transitions.

In Newtonian gravity, the structure of a spherically symmetric, self-gravitating fluid is governed by the mass continuity equation
\begin{equation}
    \frac{dM}{dr} = 4\pi r^2 \varepsilon(r),
\end{equation}
and the hydrostatic equilibrium condition
\begin{equation}
    \frac{1}{\varepsilon}\frac{dp}{dr} = -\frac{M}{r^2}.
    \label{eq:cont1}
\end{equation}
Combining these, one obtains a generalized Poisson equation for the density:
\begin{equation}
    \frac{1}{r^2}\frac{d}{dr}\left(\frac{r^2}{\varepsilon}\frac{dp}{dr}\right)
    = -4\pi \varepsilon.
\end{equation}

For the core EoS~\eqref{eq:sist1}, the derivative $dp/dr$ can be written as
\begin{align}
\frac{dp}{dr} &= 
\left\{
\frac{\alpha_1 \bigl[(\varepsilon_0/\varepsilon)(1+\tau_1)-\tau_1\bigr]\varepsilon^{\tau_1-1}}
     {\varepsilon_0^{\tau_1}\bigl(1-\varepsilon_0/\varepsilon\bigr)^2}
+ \frac{\beta_1\sigma_1}{\varepsilon_0^{\sigma_1}}\varepsilon^{\sigma_1-1} 
\right\}\frac{d\varepsilon}{dr},
\label{eq:dp1}
\end{align}
while for the crust EoS~\eqref{eq:sist2},
\begin{align}
\frac{dp}{dr} &= 
\left[
\frac{\tau_2\alpha_2}{\varepsilon}\left(\frac{\varepsilon}{\varepsilon_0}\right)^{\tau_2}
+ \frac{\sigma_2\beta_2}{\varepsilon}\left(\frac{\varepsilon}{\varepsilon_0}\right)^{\sigma_2}
\right]\frac{d\varepsilon}{dr}.
\label{eq:dp2}
\end{align}

Substituting Eq.~\eqref{eq:dp1} into Eq.~\eqref{eq:cont1} yields, for the core ($\varepsilon_0<\varepsilon$),
\begin{equation}
    \frac{1}{r^2}\frac{d}{dr}\left[
    r^2\left\{
    \frac{\alpha_1 \bigl[(\varepsilon_0/\varepsilon)(1+\tau_1)-\tau_1\bigr]}
         {\varepsilon_0^{\tau_1}\bigl(1-\varepsilon_0/\varepsilon\bigr)^2}\varepsilon^{\tau_1-2}
    + \frac{\beta_1 \sigma_1}{\varepsilon_0^{\sigma_1}}\varepsilon^{\sigma_1-2}
    \right\}\frac{d\varepsilon}{dr}\right]
    = -4\pi\varepsilon.
    \label{eq:dif1}
\end{equation}
For the crust ($\varepsilon\le\varepsilon_0$), using Eq.~\eqref{eq:dp2}, we obtain
\begin{equation}
\frac{1}{r^2}\frac{d}{dr}\left[
r^2\left(
\frac{\tau_2\alpha_2}{\varepsilon_0^{\tau_2}}\varepsilon^{\tau_2-2}
+ \frac{\sigma_2\beta_2}{\varepsilon_0^{\sigma_2}}\varepsilon^{\sigma_2-2}
\right)\frac{d\varepsilon}{dr}\right]
= -4\pi\varepsilon.
\label{eq:dif2}
\end{equation}

Equations \eqref{eq:dif1} and \eqref{eq:dif2} are nonlinear, radially symmetric elliptic equations. Their general solutions are highly non-trivial; therefore, we focus on two limiting regimes for the core:

\begin{itemize}
    \item[(i)] near the center $r=0$, where $\varepsilon_0/\varepsilon\ll1$;
    \item[(ii)] near the interface $r_1$, where $\varepsilon/\varepsilon_0\approx1$.
\end{itemize}

\subsection{Near the center: $\varepsilon_0/\varepsilon <\!\!< 1$}
\label{subsec:core_center}

For $\varepsilon_0/\varepsilon<\!\!< 1$, the first term in braces in Eq.~\eqref{eq:dif1} simplifies to
\begin{equation}
   \frac{\alpha_1 \bigl[(\varepsilon_0/\varepsilon)(1+\tau_1)-\tau_1\bigr]}
        {\varepsilon_0^{\tau_1}\bigl(1-\varepsilon_0/\varepsilon\bigr)^2}\varepsilon^{\tau_1-2}
   \approx -\frac{\tau_1\alpha_1}{\varepsilon_0^{\tau_1}}\varepsilon^{\tau_1-2},
\end{equation}
and Eq.~\eqref{eq:dif1} becomes
\begin{equation}
    \frac{1}{r^2}\frac{d}{dr}\left[
      r^2\left(
      -\frac{\tau_1\alpha_1}{\varepsilon_0^{\tau_1}}\varepsilon^{\tau_1-2}
      + \frac{\beta_1 \sigma_1}{\varepsilon_0^{\sigma_1}}\varepsilon^{\sigma_1-2}
      \right)\frac{d\varepsilon}{dr}\right]
    = -4\pi\varepsilon.
    \label{eq:dif3}
\end{equation}
The structure of Eq.~\eqref{eq:dif3} is similar to that of the crust equation~\eqref{eq:dif2}. This suggests that, at a purely differential level, the density profile near the center exhibits Lane–Emden–type behavior analogous to that near the crust, even though the underlying microphysical regimes are very different (nuclear matter vs. neutron-rich crystalline lattice with degenerate electrons). In other words, the \emph{functional form} of the equilibrium equations is similar, but the interpretation of $\varepsilon$ and the dominant contributions to $p(\varepsilon)$ differ.

\subsection{Near the interface: $\varepsilon/\varepsilon_0\approx 1$}
\label{subsec:interface_nonrel}

In the transition layer around $r=r_1$, we have $\varepsilon/\varepsilon_0\approx1$, and the factor $(1-\varepsilon_0/\varepsilon)^{-1}$ in Eq.~\eqref{eq:dif1} cannot be reliably expanded into a Taylor series. Instead, we exploit the approximate matching condition \eqref{eq:border}. Rearranging Eq.~\eqref{eq:border}, we write
\begin{equation}
   \frac{\alpha_1}{(\varepsilon_0/\varepsilon_+-1)}
   \left(\frac{\varepsilon_+}{\varepsilon_0}\right)^{\tau_1}
   \approx -\beta_1\left(\frac{\varepsilon_+}{\varepsilon_0}\right)^{\sigma_1}
   + \left[
      \alpha_2\left(\frac{\varepsilon_-}{\varepsilon_0}\right)^{\tau_2}
      + \beta_2\left(\frac{\varepsilon_-}{\varepsilon_0}\right)^{\sigma_2}
     \right]
   + \delta.
   \label{eq:border1}
\end{equation}

From the first term in brackets in Eq.~\eqref{eq:dif1}, evaluated at $\varepsilon=\varepsilon_+$, we find
\begin{equation}
    \frac{\alpha_1 \bigl[(\varepsilon_0/\varepsilon_+)(1+\tau_1)-\tau_1\bigr]}
         {\varepsilon_0^{\tau_1}\bigl(1-\varepsilon_0/\varepsilon_+\bigr)^2}\varepsilon_+^{\tau_1-2}
    = \left[
    \frac{(\varepsilon_0/\varepsilon_+)(1+\tau_1)-\tau_1}
         {\varepsilon_+^2(1-\varepsilon_0/\varepsilon_+)}
    \right]
    \frac{\alpha_1}{(1-\varepsilon_0/\varepsilon_+)}
    \left(\frac{\varepsilon_+}{\varepsilon_0}\right)^{\tau_1}.
    \label{eq:border2}
\end{equation}
Assuming that an analogous relation holds for $|\varepsilon_0/\varepsilon-1|\ll1$, Eq.~\eqref{eq:border2} can be approximated using Eq.~\eqref{eq:border1} as
\begin{align}
&\frac{\alpha_1 \bigl[(\varepsilon_0/\varepsilon_+)(1+\tau_1)-\tau_1\bigr]}
      {\varepsilon_0^{\tau_1}(1-\varepsilon_0/\varepsilon_+)^2}
      \varepsilon_+^{\tau_1-2}
\approx
\left[
\frac{(\varepsilon_0/\varepsilon_+)(1+\tau_1)-\tau_1}
     {\varepsilon_+^2(1-\varepsilon_0/\varepsilon_+)}
\right]\nonumber\\[0.3em]
&\hspace{1cm}\times\left\{
\beta_1\left(\frac{\varepsilon_+}{\varepsilon_0}\right)^{\sigma_1}
- \left[
   \alpha_2\left(\frac{\varepsilon_-}{\varepsilon_0}\right)^{\tau_2}
   + \beta_2\left(\frac{\varepsilon_-}{\varepsilon_0}\right)^{\sigma_2}
  \right]-\delta\right\}.
\label{eq:aproxcc}
\end{align}

This motivates the definition of an effective coupling function
\begin{equation}
\begin{aligned}
    f(\varepsilon) &=
    \left[
    \frac{(\varepsilon_0/\varepsilon_+)(1+\tau_1)-\tau_1}
         {\varepsilon_+^2(1-\varepsilon_0/\varepsilon_+)}
    \right]\\[0.2em]
    &\quad\times\left\{
    \beta_1\left(\frac{\varepsilon_+}{\varepsilon_0}\right)^{\sigma_1}
    - \left[
       \alpha_2\left(\frac{\varepsilon_-}{\varepsilon_0}\right)^{\tau_2}
       + \beta_2\left(\frac{\varepsilon_-}{\varepsilon_0}\right)^{\sigma_2}
      \right] - \delta
    \right\}.
\end{aligned}
\end{equation}
The function $f(\varepsilon)$ encodes how the core and crust EoS couple in the transition layer. It effectively measures the correction to the local pressure-differential operator arising from the mismatch between the two branches.

Introducing the dimensionless energy density parameter $\eta=\eta(r)$ via $\varepsilon_+ = \eta \varepsilon_0$ ($\eta>1$) and $\varepsilon_-=\varepsilon_0/\eta$, the coupling term becomes
\begin{equation}
\begin{aligned}
    f(\eta) &= 
    \left[
    \frac{(1/\eta)(1+\tau_1)-\tau_1}
         {(\eta\varepsilon_0)^2(1-1/\eta)}
    \right]\\[0.2em]
    &\quad\times\left\{
    \beta_1\eta^{\sigma_1}
    - \left[
        \alpha_2\eta^{-\tau_2}
        + \beta_2\eta^{-\sigma_2}
      \right] - \delta
    \right\}.
\end{aligned}
\end{equation}

Near $r_1$, Eq.~\eqref{eq:dif1} can then be recast as
\begin{equation}
    \frac{1}{r^2}\frac{d}{dr}\left[
    r^2\left(
    f(\eta) + \frac{\beta_1\sigma_1}{\varepsilon_0^{\sigma_1}}(\eta\varepsilon_0)^{\sigma_1-2}
    \right)\frac{d\eta}{dr}\right]
    = -4\pi\eta.
    \label{eq:dif1x}
\end{equation}
This equation can be interpreted as an effective field equation governing the transition zone, with $A(\eta)$ acting as a density-dependent ``diffusion'' or ``conductivity'' coefficient:
\begin{equation}
    A(\eta) = f(\eta)
    + \frac{\beta_1\sigma_1}{\varepsilon_0^{\sigma_1}}(\eta\varepsilon_0)^{\sigma_1-2},
\end{equation}
so that
\begin{equation}\label{eq:aeta}
    \frac{1}{r^2}\frac{d}{dr}\left[r^2 A(\eta)\frac{d\eta}{dr}\right] = -4\pi\eta.
\end{equation}
This admits a first-order system formulation by defining $u=\eta$ and $v = r^2 A(u)\,du/dr$:
\begin{equation}
    \label{eq:sistema1}
    \begin{cases}
        u' = \dfrac{v}{r^2 A(u)}, \\[0.4em]
        v' = -4\pi r^2 u.
    \end{cases}
\end{equation}
The evolution of $u(r)$ in the transition layer is thus governed by a coupled system resembling a generalized Lane–Emden equation with a variable, density-dependent diffusion coefficient.

\section{Relativistic Treatment}
\label{sec:rel}

We now consider the full relativistic description of the star via the TOV equations. For a static, spherically symmetric, non-charged NS, the TOV system reads
\begin{align}
    \frac{dp_r}{dr} &= -\frac{(\varepsilon_r+p_r)(M_r+4\pi r^3 p_r)}
    {r^2\bigl(1-2M_r/r\bigr)},
    \label{eq:TOV_p}\\
    \frac{dM_r}{dr} &= 4\pi r^2 \varepsilon_r,
    \label{eq:TOV_M}
\end{align}
with boundary conditions $p_r(0)=p_f$ and $M_r(0)=0$. We work with reduced quantities, defining the dimensionless energy density and mass as
\begin{equation}
    \varepsilon_r = \frac{\varepsilon}{\varepsilon_0}, \qquad
    M_r = \frac{M}{M_\odot},
\end{equation}
so that $M_r$ is the mass in units of solar masses.

Analytical solutions of Eqs.~\eqref{eq:TOV_p}–\eqref{eq:TOV_M} are scarce when realistic, nonlinear EoS are used. In our case, the piecewise EoS \eqref{eq:sistdimensionless1}–\eqref{eq:sistdimensionless2} leads to a strong sensitivity to initial conditions and parameter values. Thus, we rely on a numerical search over the parameter space constrained by the thermodynamic conditions discussed earlier and by the different $(\tau_1,\sigma_1)$ regimes in Table~\ref{tab:regimes}.

\subsection{Parameter regimes and mass–radius relations}

We are looking for values that maintain causality, $dp/d\varepsilon> 0$, and that allow sufficient curvature to accelerate the flattening of $\mu(\varepsilon_0)$. Moreover, we explore the parameter space by selecting ranges for $(\alpha_1,\beta_1,\tau_1,\sigma_1)$ in the core and $(\alpha_2,\beta_2,\tau_2,\sigma_2)$ in the crust, subject to the stability and causality constraints. For each regime specified in Table~\ref{tab:regimes}, we adjust parameters to produce NS configurations with realistic densities and masses, including low-mass configurations.

Table~\ref{tab:tabparam1} displays representative parameter sets and resulting stellar properties for regimes where $\tau_1\leq -1$ and $0\leq\sigma_1$, corresponding to the first line of Table~\ref{tab:regimes}. These sets produce stable configurations with masses ranging from $0.99$ to $2.05\,M_\odot$.

\begin{table}[t]
    \centering
    \begin{tabular}{|c|c|c|c|c|c|c|c|c|c|c|c|}
    \hline
Regime & $M$ ($M_{\odot}$) & $R$ (km) & $\varepsilon_c/\varepsilon_0$ & $\alpha_1$ & $\beta_1$ & $\tau_1$ & $\sigma_1$  & $\alpha_2$ & $\beta_2$ & $\tau_2$ & $\sigma_2$  \\
\hline
\multirow{3}{3.0cm}{$\tau_1\leq -1,\,0\leq \sigma_1$}  
&0.79&11.30&95.46&1.75&0.25&-1.89&2.50&2.88&2.63&0.80&3.75 \\
&1.78&10.01&95.46&1.75&0.25&-1.89&2.50&2.88&2.63&0.98&3.85 \\
&1.94&11.28&23.78&-0.37&0.17&-1.00&2.50&2.88&2.63&0.98&3.90 \\
\hline
    \end{tabular}
    \caption{Representative parameter sets and resulting NS properties for regimes with $\tau_1\leq -1$ and $0\leq\sigma_1$ (first line of Table~\ref{tab:regimes}). The corresponding mass–radius curves are shown in Fig.~\ref{fig:tov1}.}
    \label{tab:tabparam1}
\end{table}

\begin{figure}[t]
   \centering
   \includegraphics[width=7cm,height=5cm]{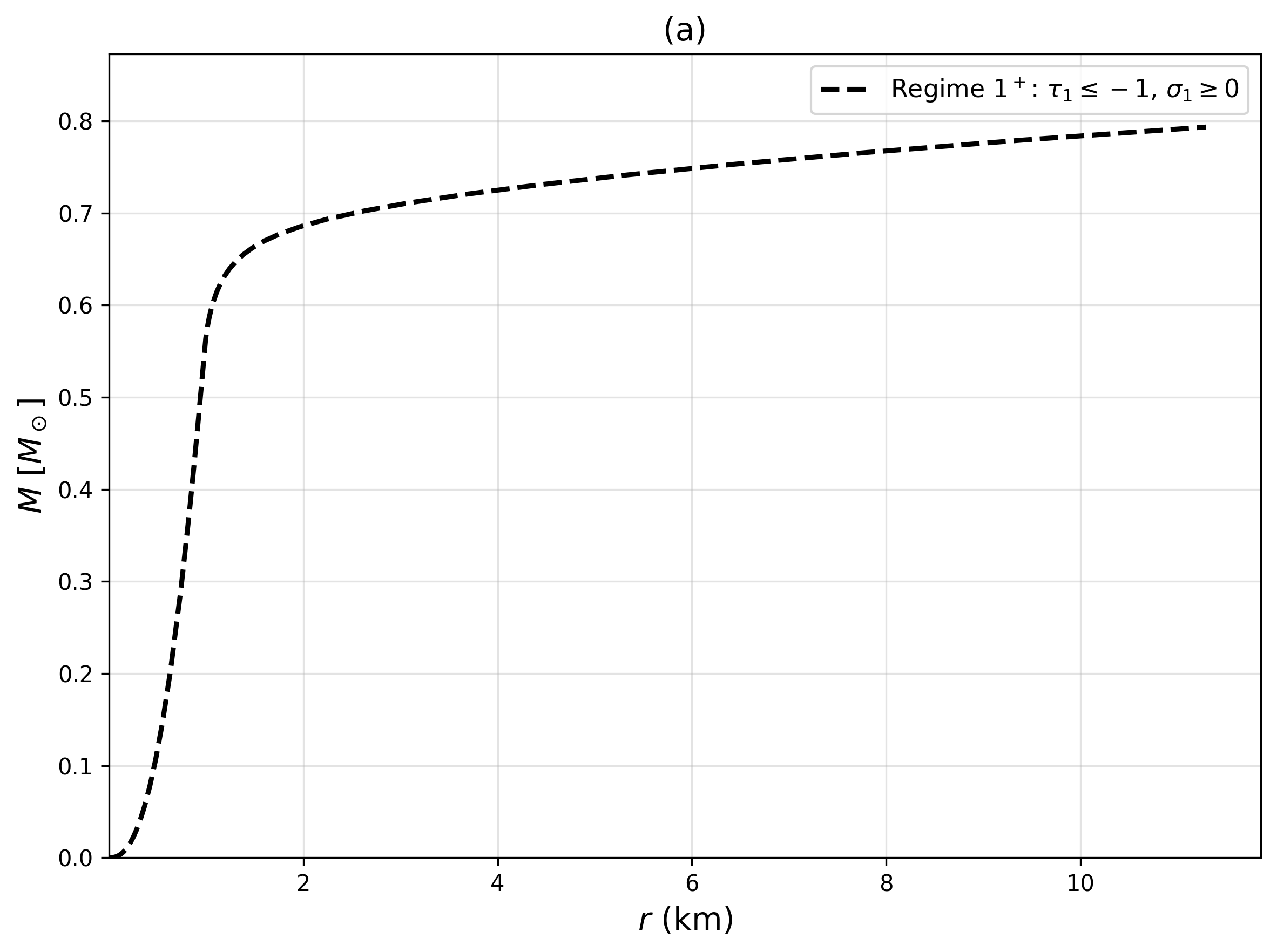}
   \includegraphics[width=7cm,height=5cm]{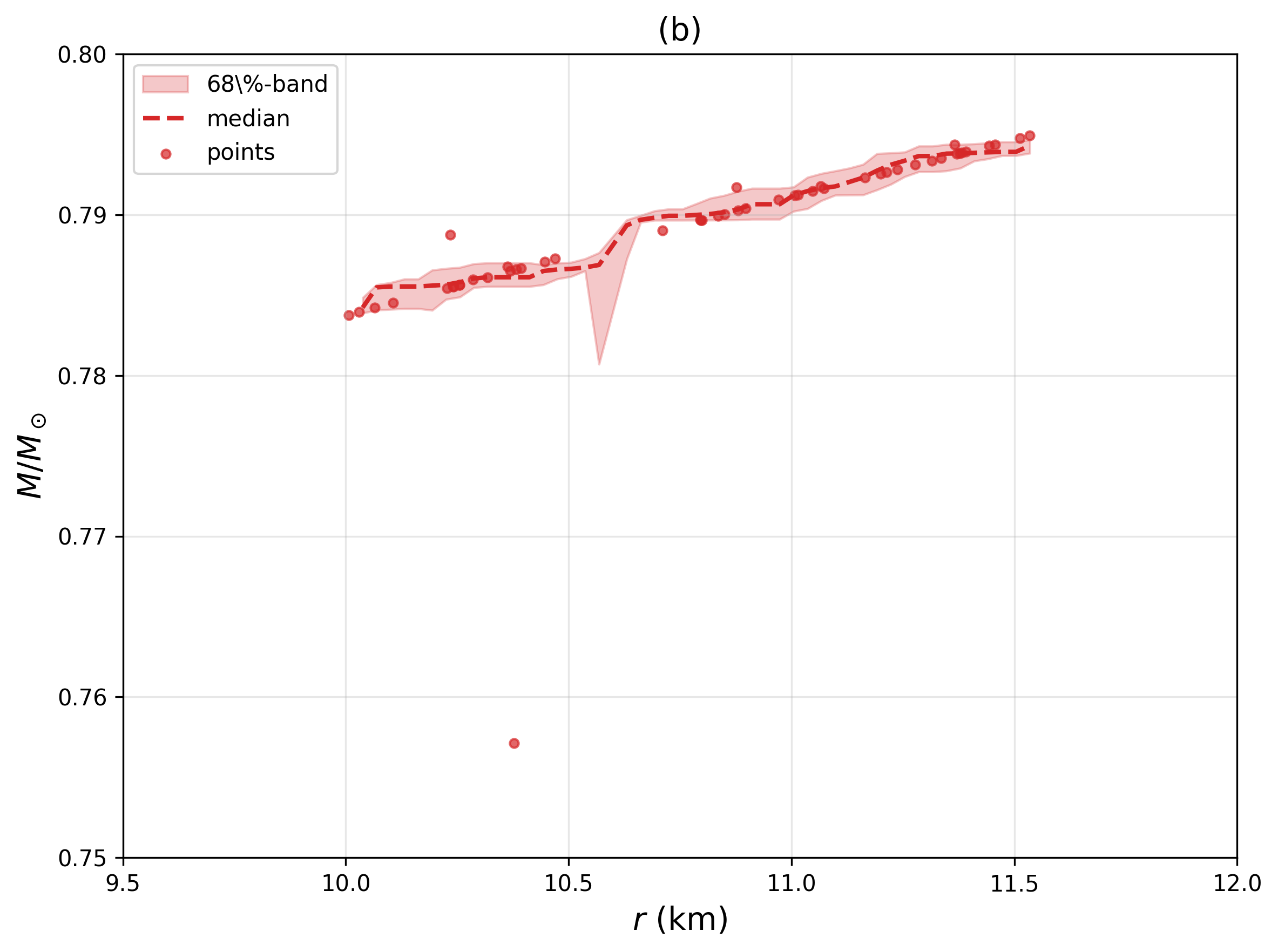}
   \includegraphics[width=7cm,height=5cm]{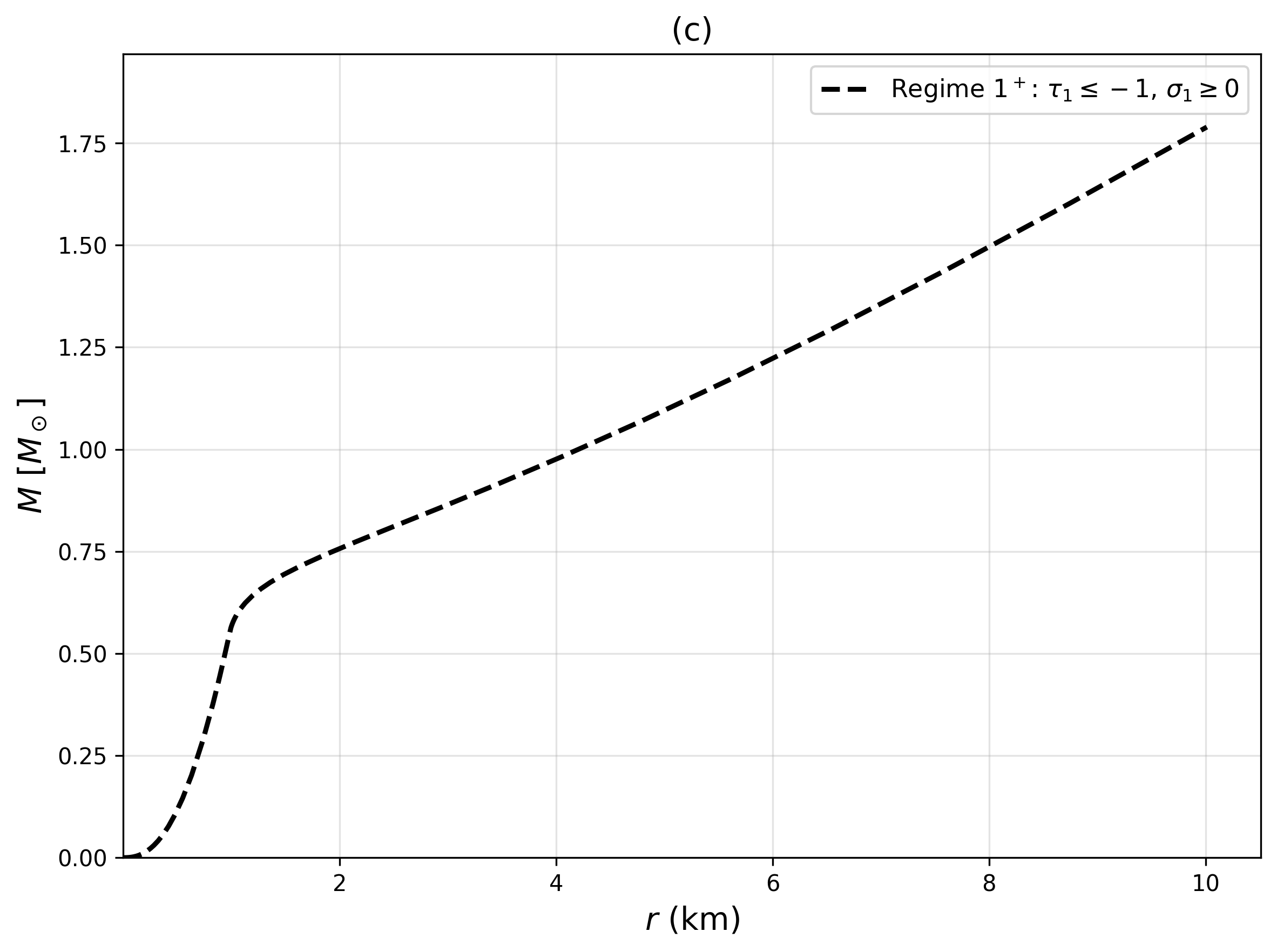}
   \includegraphics[width=7cm,height=5cm]{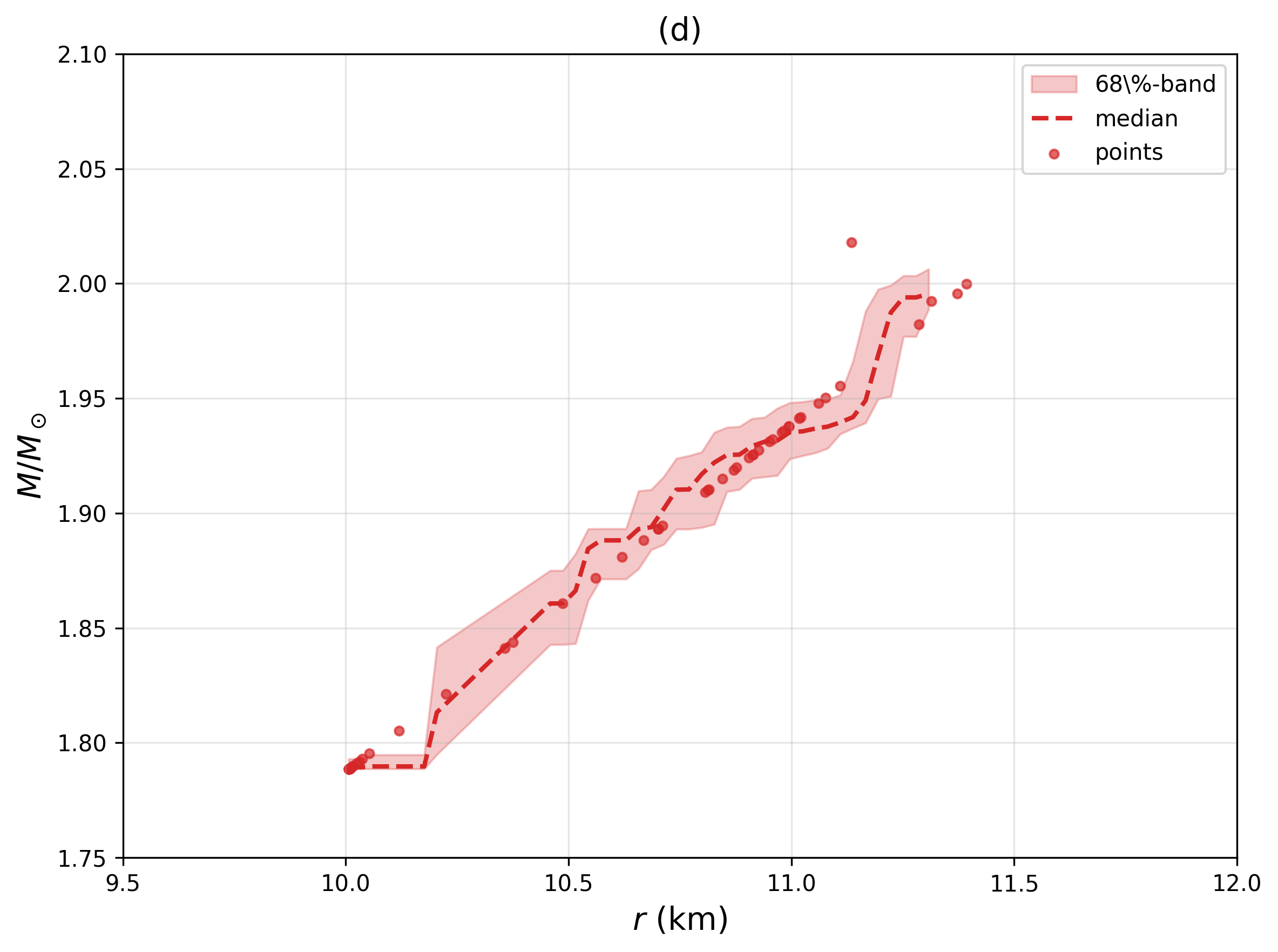}
   \includegraphics[width=7cm,height=5cm]{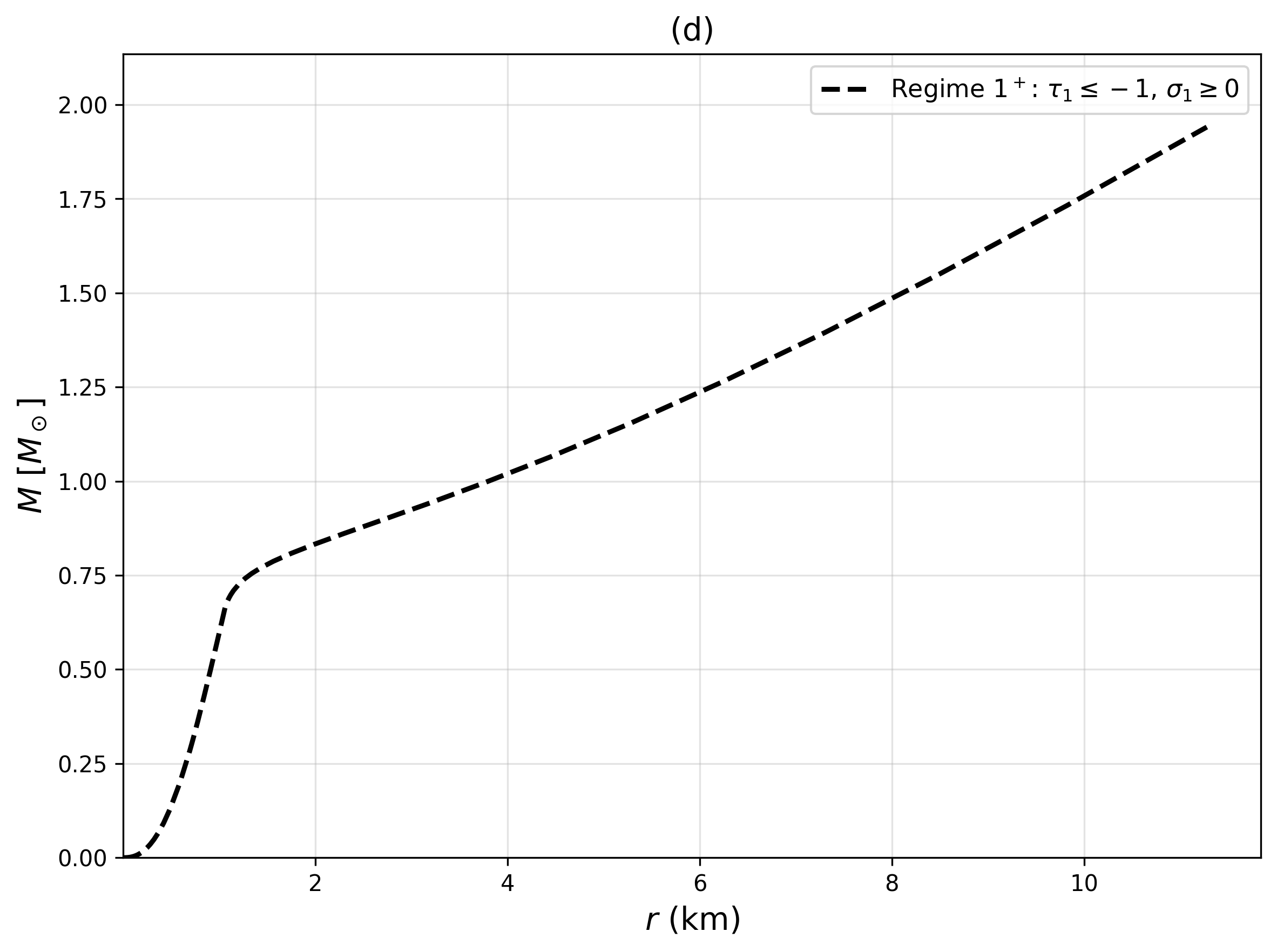}
   \includegraphics[width=7cm,height=5cm]{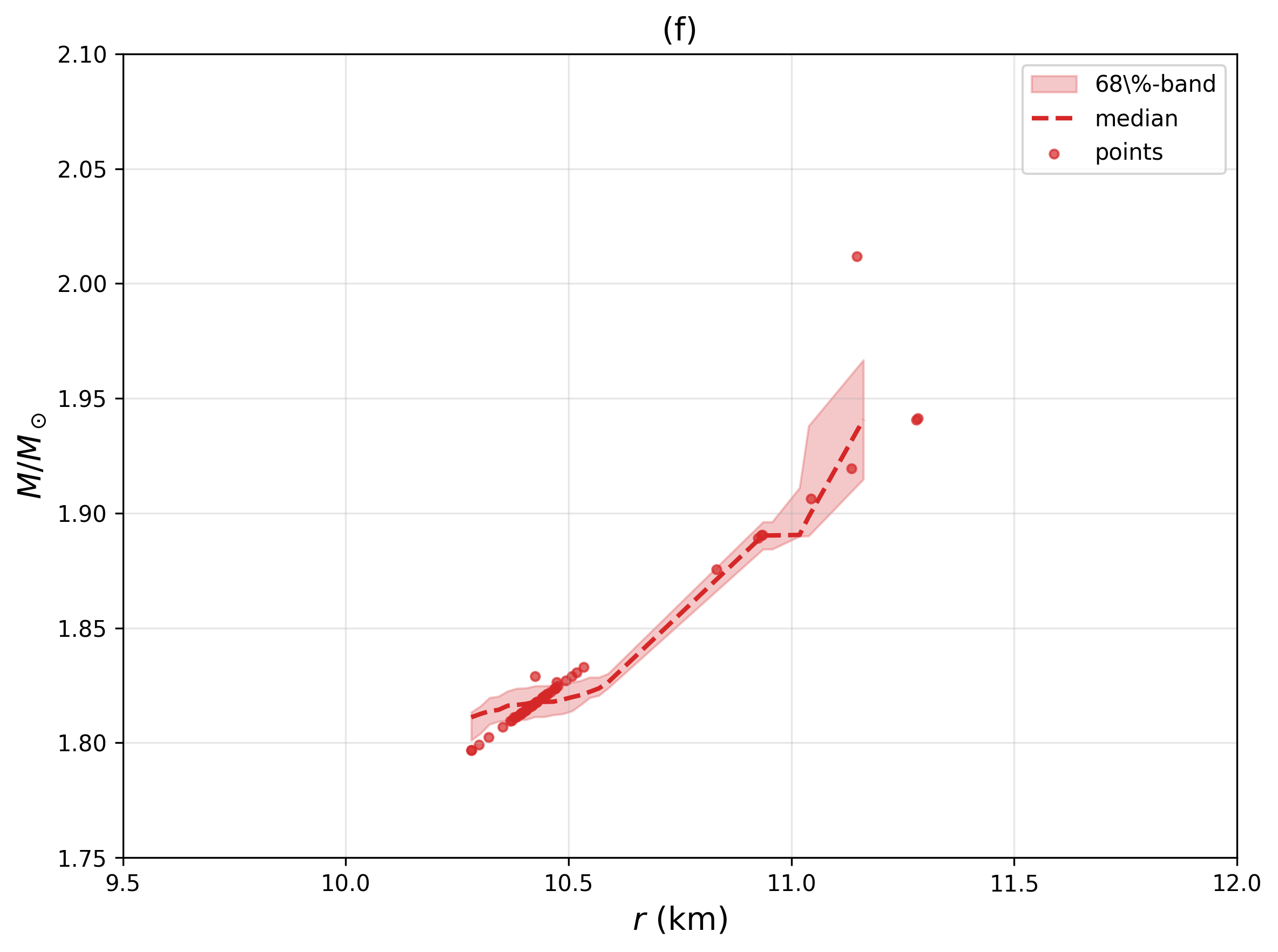}
   \caption{Mass–radius relations from the TOV equations for the parameter sets in Table~\ref{tab:tabparam1}. Panels (a), (c), and (e): $M(r)$ profiles for the configurations in each regime. Panels (b), (d), and (f): families of $M/M_\odot$ vs.\ $R$ curves obtained by scanning over central densities $\varepsilon_c$.}
    \label{fig:tov1}
\end{figure}

Figure~\ref{fig:tov1} displays mass–radius relations derived from the TOV equations using the parameter sets in Table~\ref{tab:tabparam1}. Panels (a), (c), and (e) display the $M(r)$ profiles for individual configurations. In contrast, panels (b), (d), and (f) display the $M/M_\odot$ vs.\ $R$ curves obtained by varying the central density $\varepsilon_c$ across a range of values, highlighting how each parameterization generates a family of equilibrium configurations.

Additional regimes in Table~\ref{tab:regimes} correspond to $0<\tau_1$ with either $\sigma_1<0$ or $\sigma_1>0$, as well as mixed-sign combinations. Table~\ref{tab:tabparam2} lists the parameter sets for two such regimes, together with the resulting NS properties.

\begin{table}[t]
    \centering
    \begin{tabular}{|c|c|c|c|c|c|c|c|c|c|c|c|}
    \hline
Regime & $M$ ($M_{\odot}$) & $R$ (km) & $\varepsilon_c/\varepsilon_0$ & $\alpha_1$ & $\beta_1$ & $\tau_1$ & $\sigma_1$  & $\alpha_2$ & $\beta_2$ & $\tau_2$ & $\sigma_2$  \\
\hline
\multirow{3}{2.6cm}{$0<\tau_1,\,\sigma_1<0$}
&1.00&10.69&56.44&-0.42&2.01&1.50&-2.50&1.65&0.68&0.75&2.05 \\
&1.21&4.32&18.33&-1.37&0.25&1.67&-2.04&0.42&2.63&0.39&3.83 \\
&2.16&10.16&51.00&0.70&2.21&0.01&-0.21&0.09&0&0.32&0 \\
\hline
\multirow{3}{2.6cm}{$0<\tau_1,\,0\leq\sigma_1$}
&1.00&10.40&78.22&-0.41&1.91&3.25&0.67&1.72&0.68&0.75&78.22 \\
&1.25&3.91&97.28&-0.51&0.04&3.03&2.58&0.42&2.63&0.39&3.83 \\
&2.02&10.16&23.78&1.07&0.30&0.02&1.67&0.17&0&0.45&0 \\
\hline
    \end{tabular}
    \caption{Representative parameter sets and NS properties for regimes with $0<\tau_1$ and either $\sigma_1<0$ or $0\leq\sigma_1$. The corresponding mass–radius curves are shown in Fig.~\ref{fig:tov2}.}
    \label{tab:tabparam2}
\end{table}

Finally, Table~\ref{tab:tabparam3_4} focuses on the most interesting regimes from the perspective of phase-transition behavior: $0<\tau_1\lesssim0.142$ with $\sigma_1<0$ and $0.143\lesssim\tau_1\lesssim0.207$ with $\sigma_1>0$, corresponding respectively to lines 3 and 4 of Table~\ref{tab:regimes}. These regimes show pronounced competition between thermal and interaction contributions. Observe that the critical density $\varepsilon_c$ is found to lie within the range $(10–150)\times\varepsilon_0$, depending on $(\tau_1,\sigma_1)$.

\begin{table}[t]
    \centering
    \begin{tabular}{|c|c|c|c|c|c|c|c|c|c|c|c|}
    \hline
Regime & $M$ ($M_{\odot}$) & $R$ (km) & $\varepsilon_c/\varepsilon_0$ & $\alpha_1$ & $\beta_1$ & $\tau_1$ & $\sigma_1$  & $\alpha_2$ & $\beta_2$ & $\tau_2$ & $\sigma_2$  \\
\hline
\multirow{3}{4.5cm}{$0<\tau_1\lesssim 0.142,\,\sigma_1<0$} 
&0.99&4.88&95.46&-0.22&2.43&0.10&-0.01&0.19&0.19&0.25&2.21 \\ 
&1.47&10.32&21.96&0.55&2.40&0.08&-0.01&0.11&0.11&1.70&1.20 \\
&2.29&10.14&3.81&0.10&1.90&0.01&-0.49&0.65&1.20&0.64&3.29 \\
\hline
\multirow{3}{4.5cm}{$0.143\lesssim \tau_1\lesssim 0.207,\, \sigma_1>0$}
&1.00&5.22&95.46&-1.00&2.40&0.20&2.50&0.17&0.14&0.25&2.12 \\
&1.45&10.03&13.80&-0.42&2.50&0.20&2.50&0.11&0.11&1.70&1.80 \\
&2.05&10.05&20.15&-0.65&0.30&0.20&2.50&0.65&1.00&0.64&3.29 \\
\hline
    \end{tabular}
    \caption{Parameter sets for regimes with $0<\tau_1\lesssim 0.142,\;\sigma_1<0$ and $0.143\lesssim \tau_1\lesssim 0.207,\;\sigma_1>0$ (lines 3 and 4 of Table~\ref{tab:regimes}). These regimes are particularly relevant for phase-transition behavior in the core. The associated mass–radius curves are shown in Fig.~\ref{fig:tov3_4}.}
    \label{tab:tabparam3_4}
\end{table}

\begin{figure}[t]
   \centering
   \includegraphics[width=7cm,height=5cm]{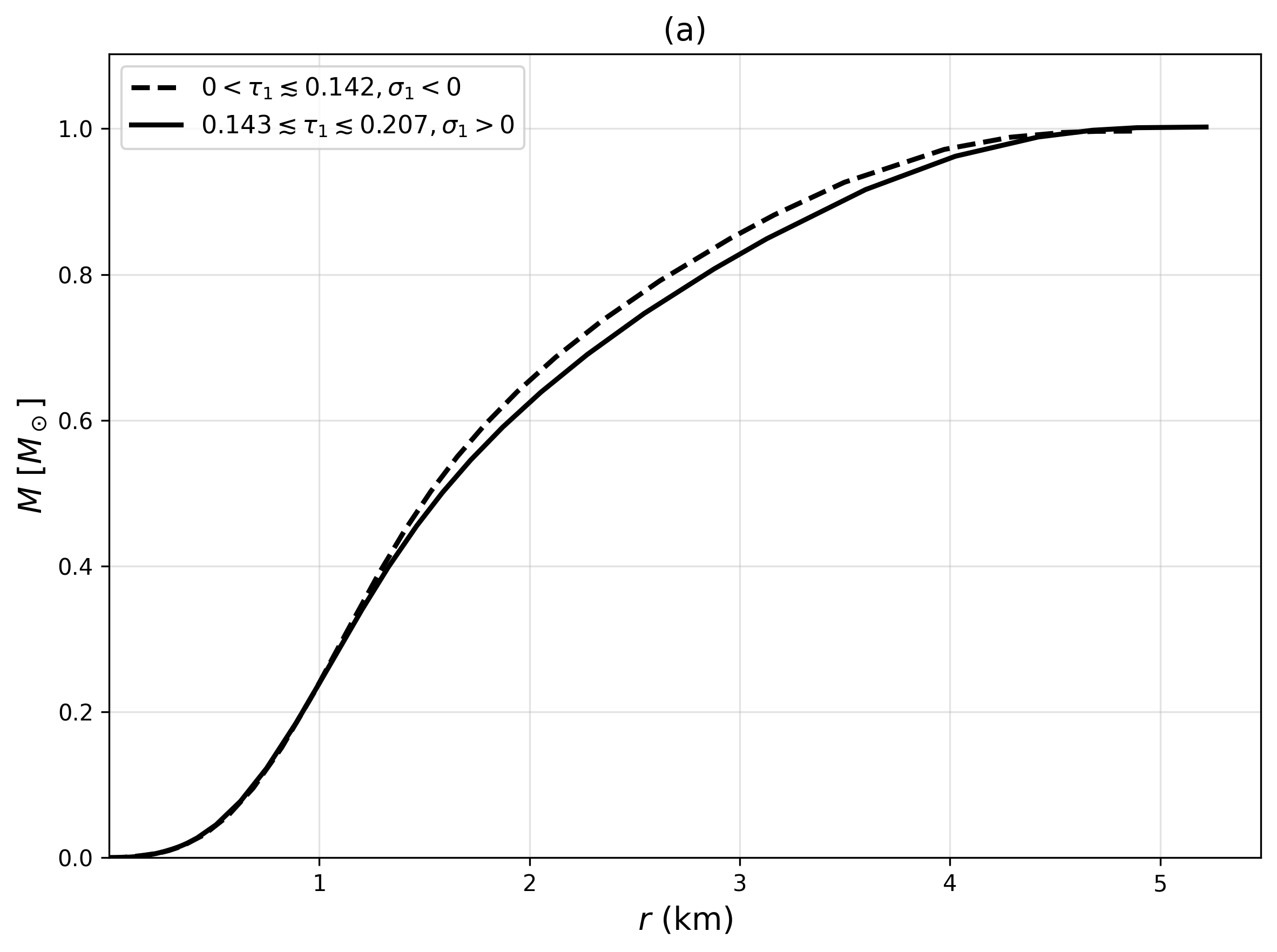}
   \includegraphics[width=7cm,height=5cm]{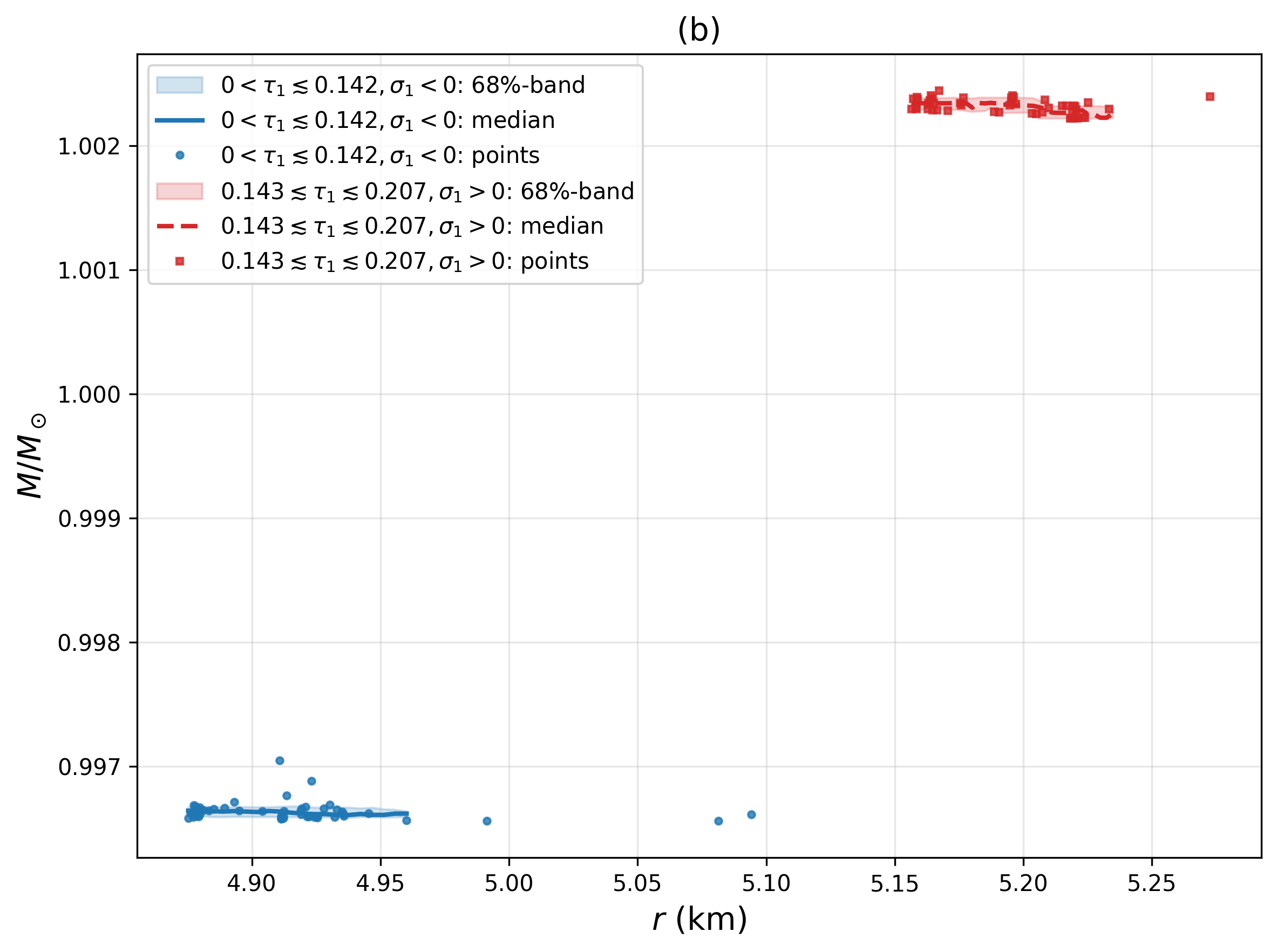}
   \includegraphics[width=7cm,height=5cm]{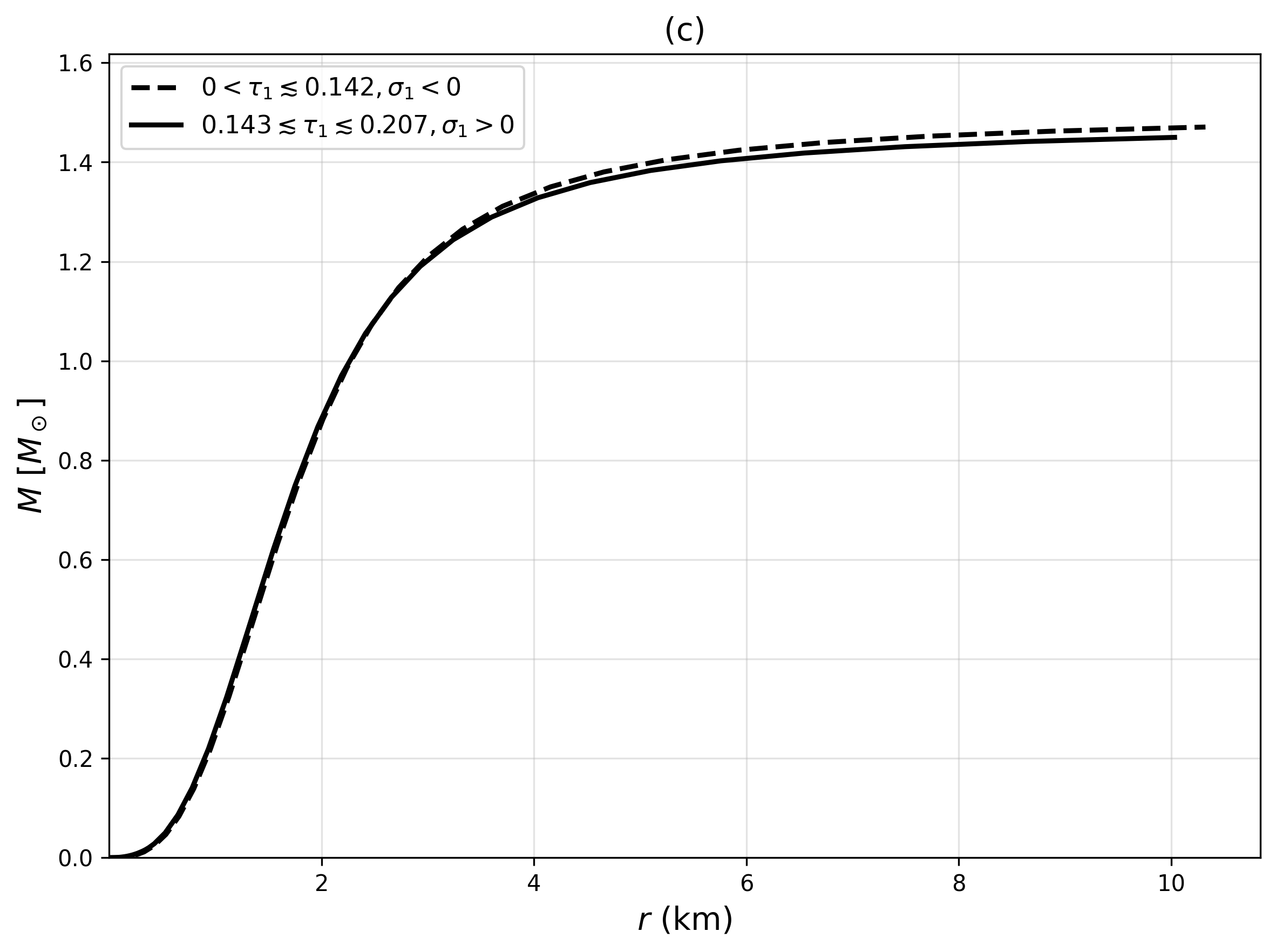}
   \includegraphics[width=7cm,height=5cm]{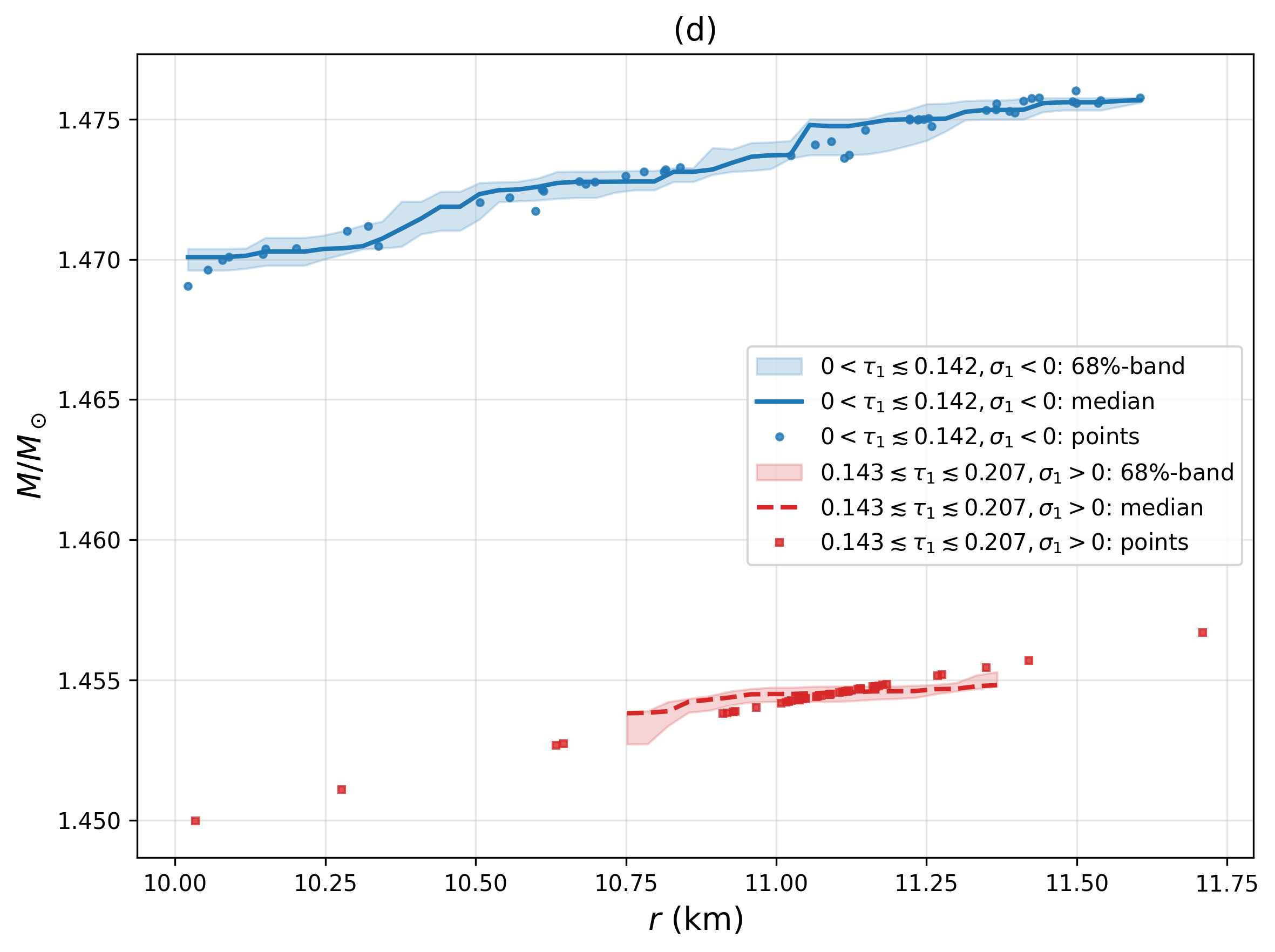}
   \includegraphics[width=7cm,height=5cm]{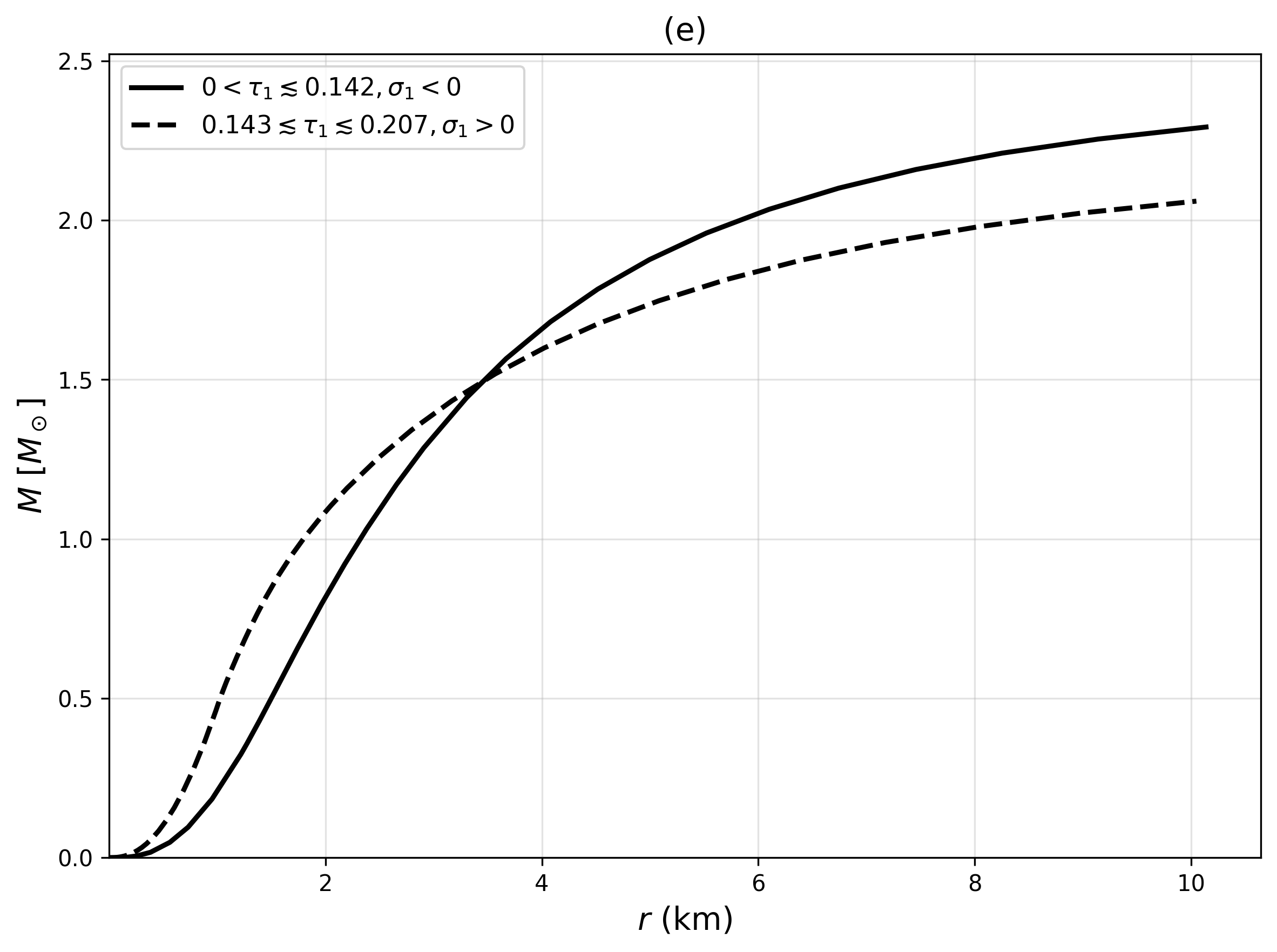}
   \includegraphics[width=7cm,height=5cm]{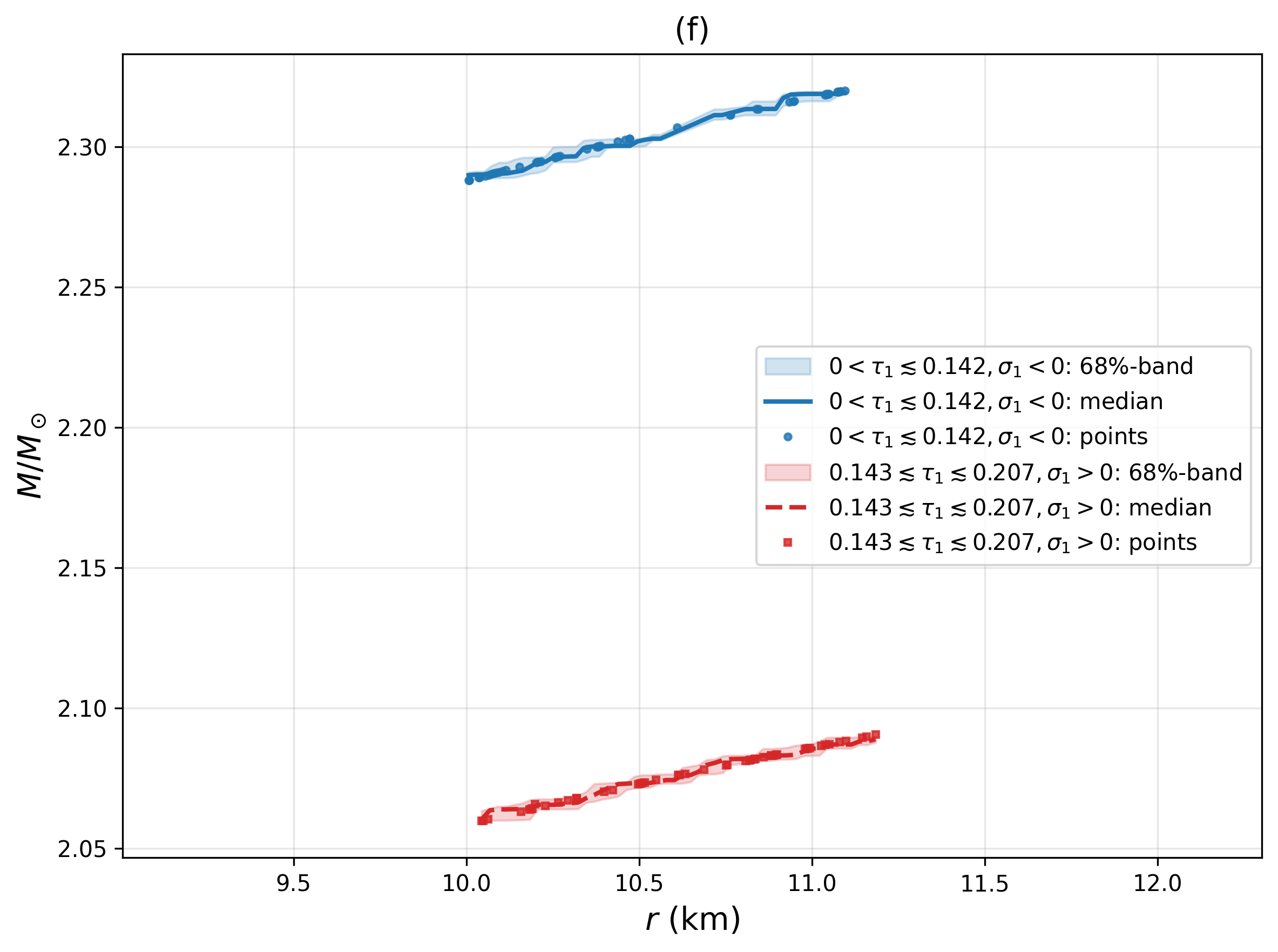}
   \caption{Mass–radius relations for the parameter sets in Table~\ref{tab:tabparam3_4}, corresponding to the most phase-transition–sensitive regimes. Panels (a), (c), and (e) show $M(r)$ for each configuration. Panels (b), (d), and (f): $M/M_\odot$ vs.\ $R$ as central densities are varied.}
    \label{fig:tov3_4}
\end{figure}

The corresponding TOV solutions are plotted in Fig.~\ref{fig:tov3_4}. Panels (a), (c), and (e) show $M(r)$ for selected masses, while panels (b), (d), and (f) illustrate families of $M/M_\odot$ vs.\ $R$ curves generated by scanning over central densities. These regimes are also used in the interface analysis (Fig.~\ref{fig:lane_emden_mod}) and in the discussion of chemical potentials in the next subsection.

\begin{figure}[!h]
   \centering
   \includegraphics[width=7cm,height=5cm]{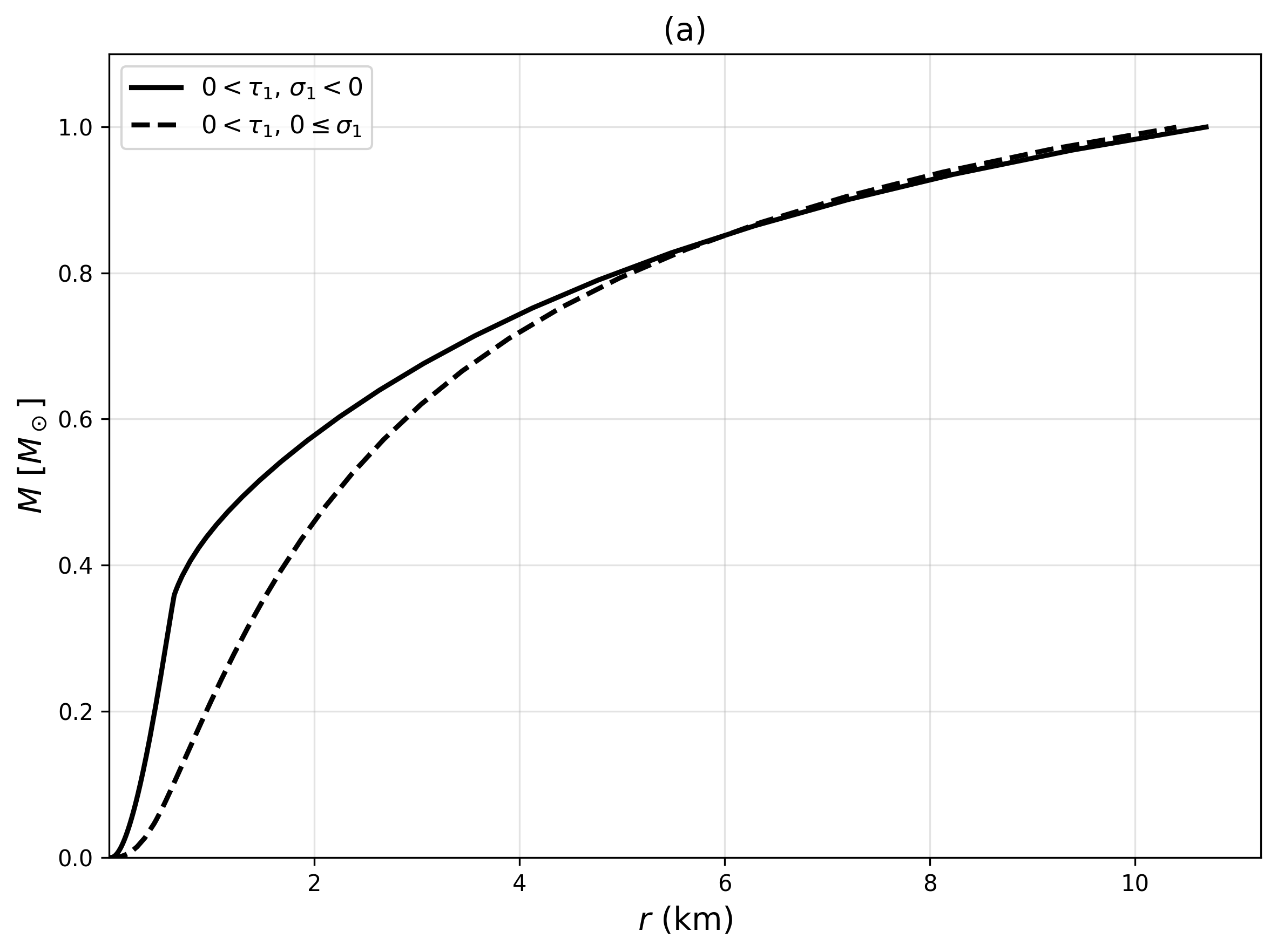}
   \includegraphics[width=7cm,height=5cm]{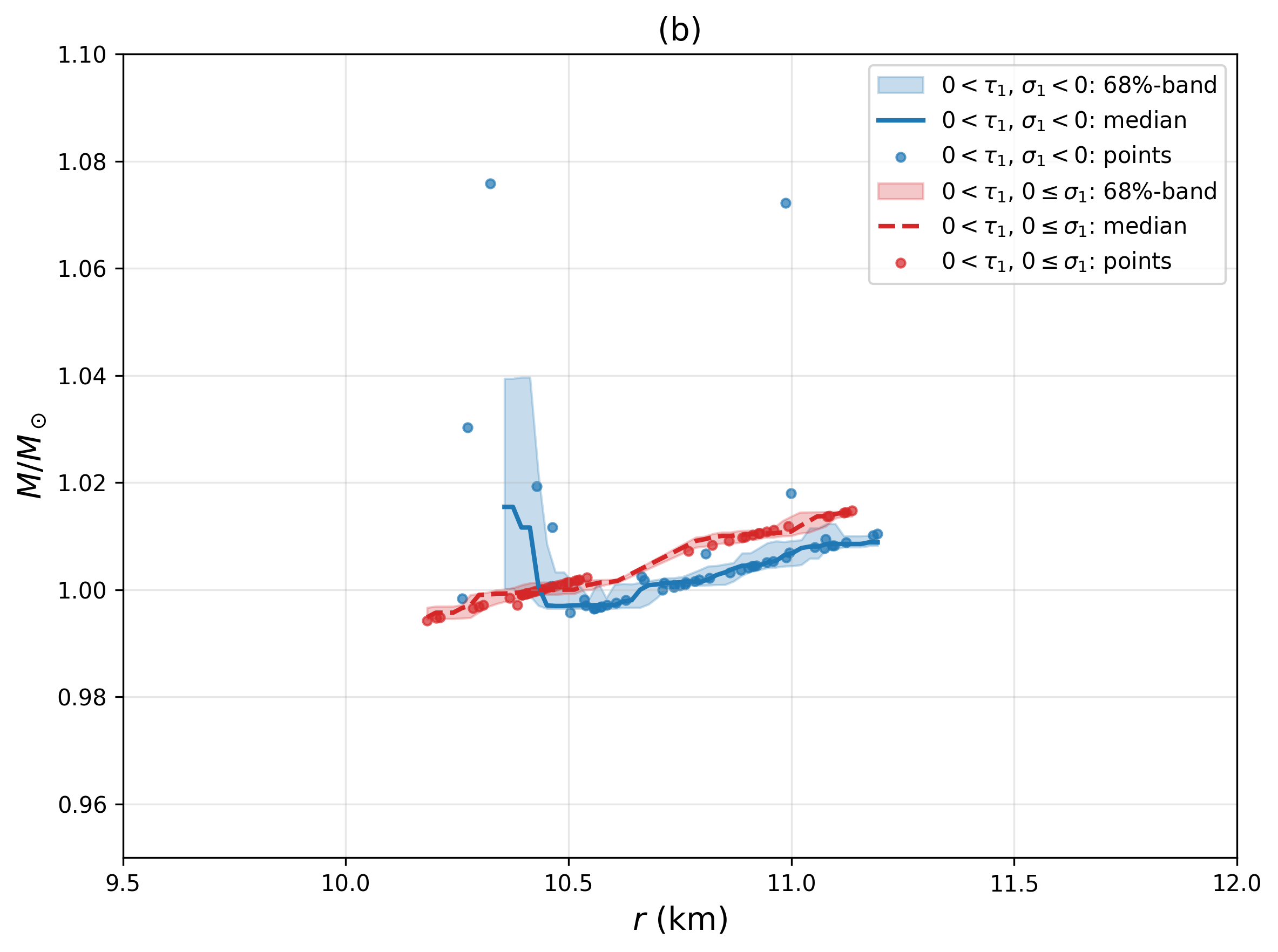}
   \includegraphics[width=7cm,height=5cm]{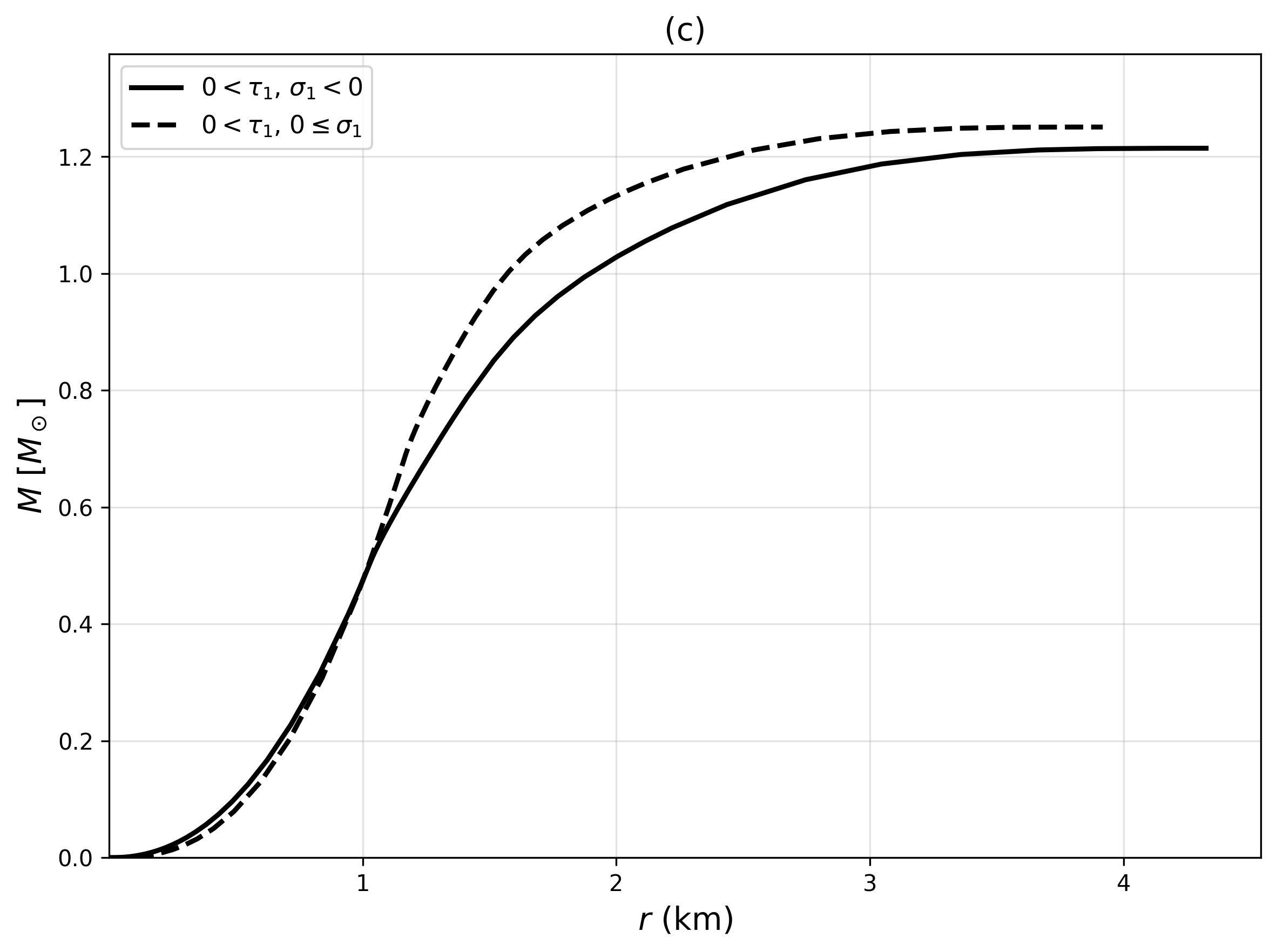}
   \includegraphics[width=7cm,height=5cm]{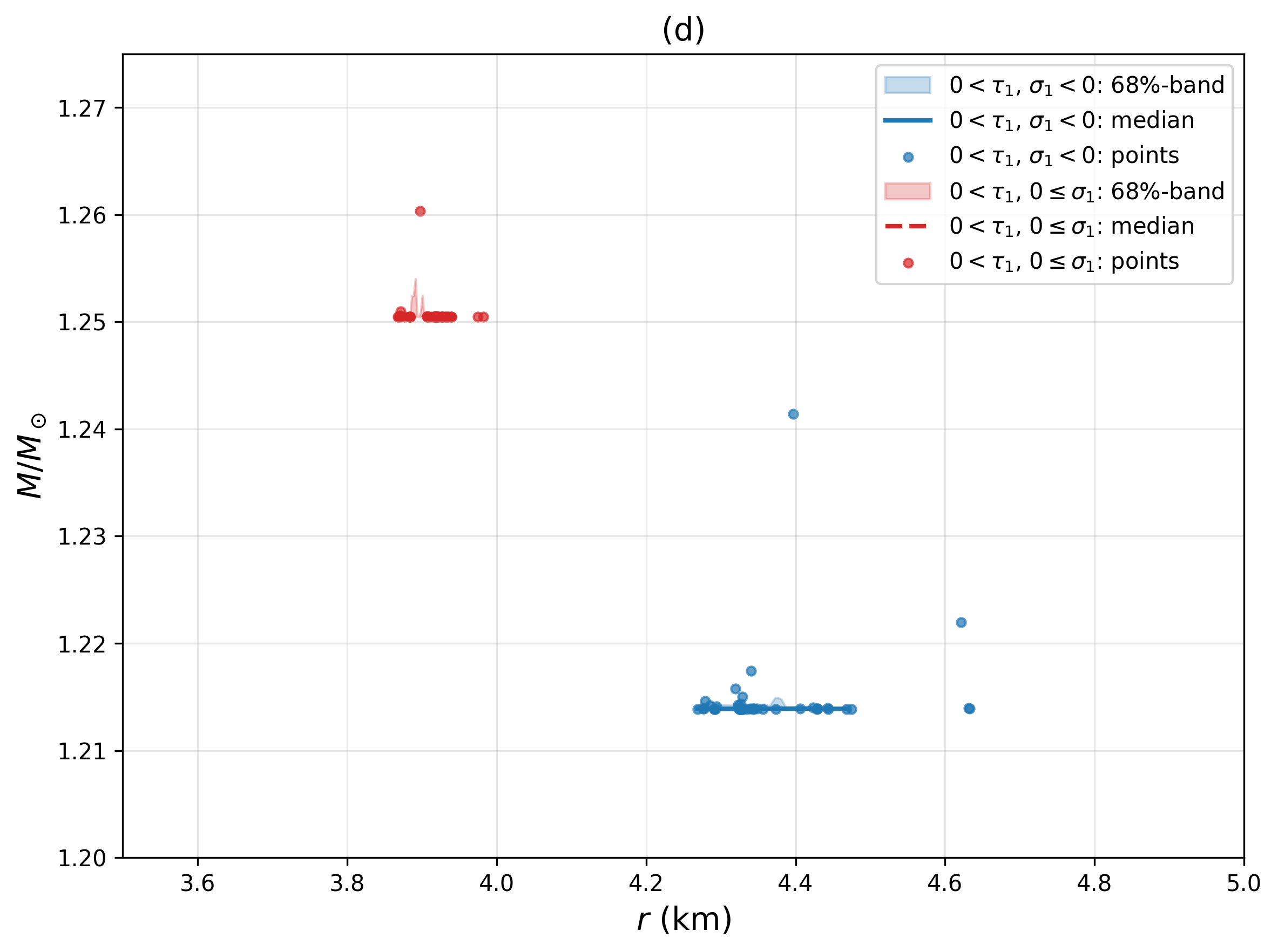}
   \includegraphics[width=7cm,height=5cm]{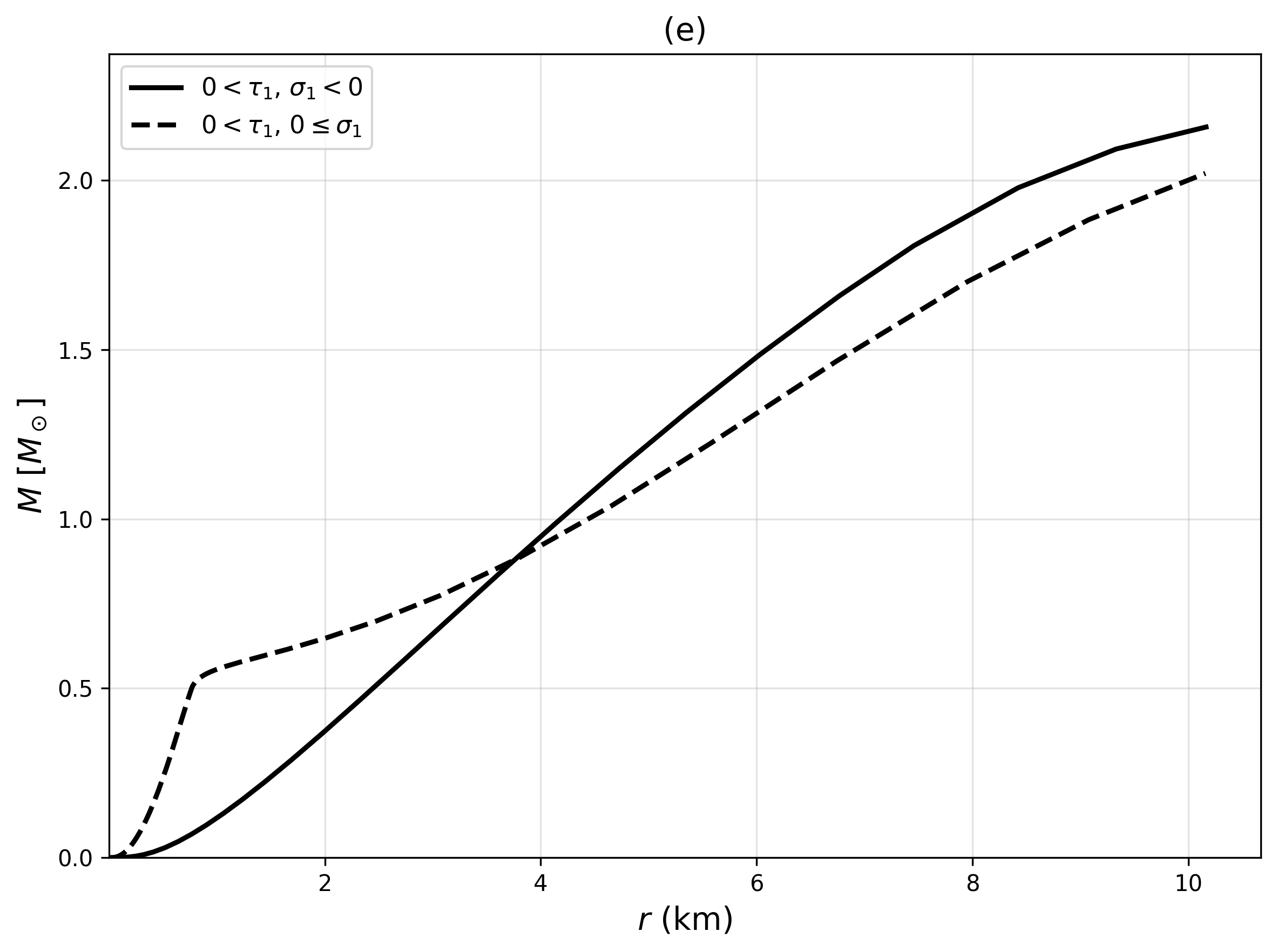}
   \includegraphics[width=7cm,height=5cm]{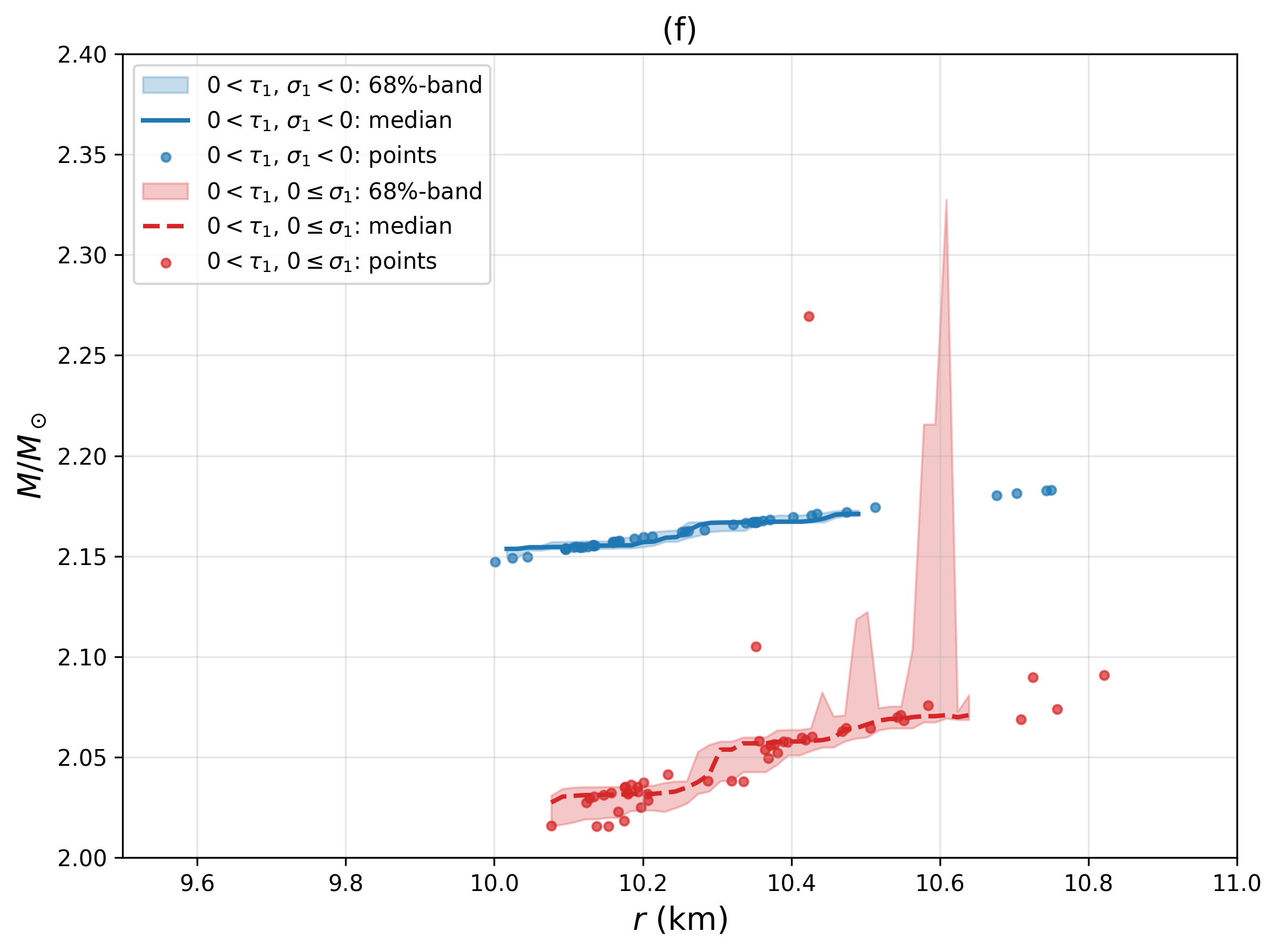}
   \caption{Mass–radius relations for the parameter sets of Table~\ref{tab:tabparam2}. Panels (a), (c), and (e): $M(r)$ for selected configurations in each regime. Panels (b), (d), and (f): $M/M_\odot$ vs.\ $R$ for several central densities, highlighting how the choice of $(\tau_1,\sigma_1)$ affects the stiffness of the EoS and the maximum mass. Observe that the increase in $\sigma_1$ turns the EoS more rigid and displaces the maximum amount of mass.}
    \label{fig:tov2}
\end{figure}

Figure~\ref{fig:tov2} shows the mass–radius behavior for these regimes. Again, panels (a), (c), and (e) show $M(r)$ for each configuration; panels (b), (d), and (f) display the associated families of $M(M_\odot)$ vs.\ $R$ curves obtained by sampling the central density. These regimes illustrate how different combinations of $(\tau_1,\sigma_1)$ interpolate between softer and stiffer behaviors at high density, altering the maximum mass and the shape of the $M$–$R$ curves.

\subsection{Phase-transition signatures and chemical potential}

\begin{figure}[t]
   \centering
   \includegraphics[width=7.5cm,height=9.5cm]{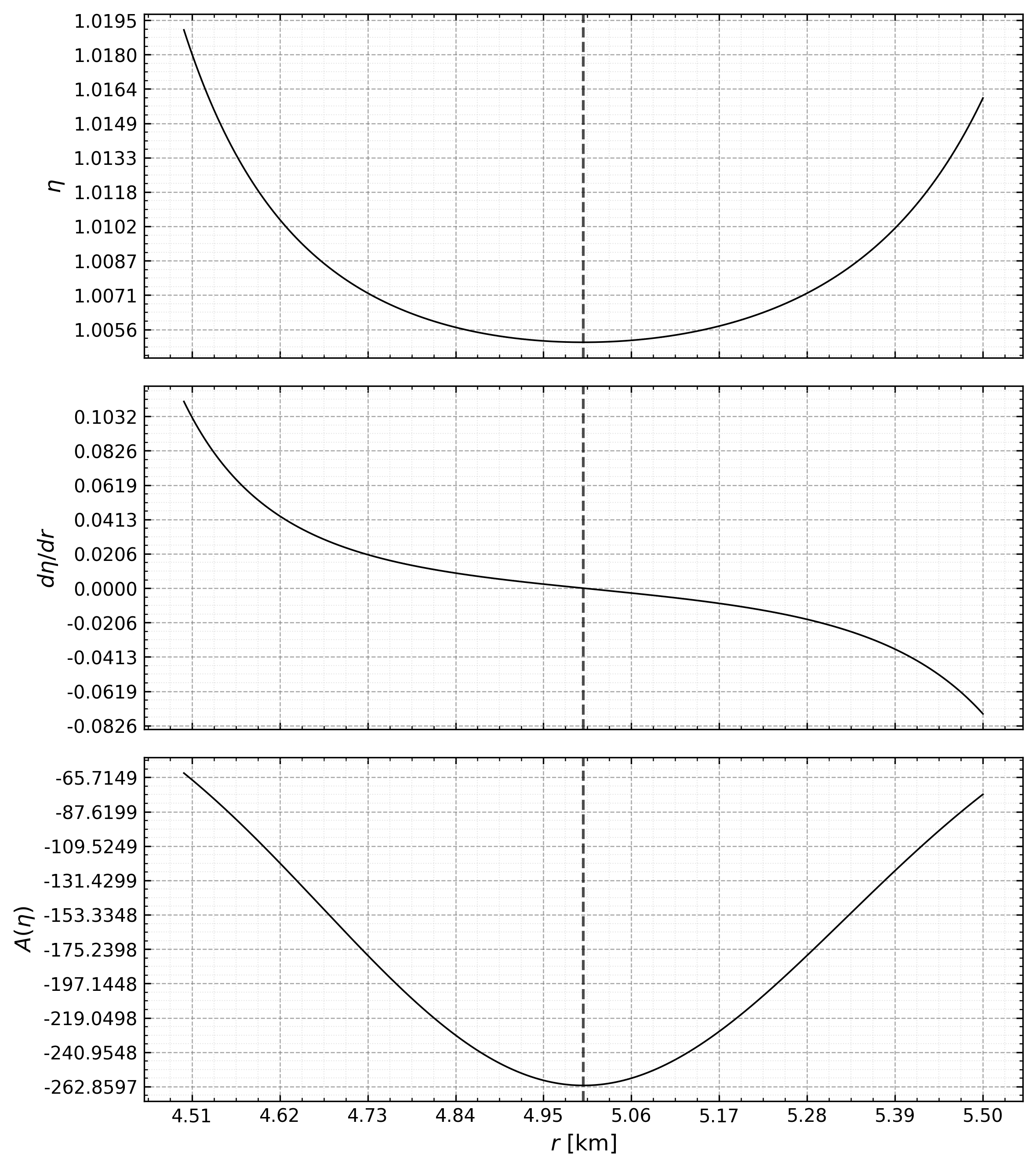}
   \includegraphics[width=7.5cm,height=9.5cm]{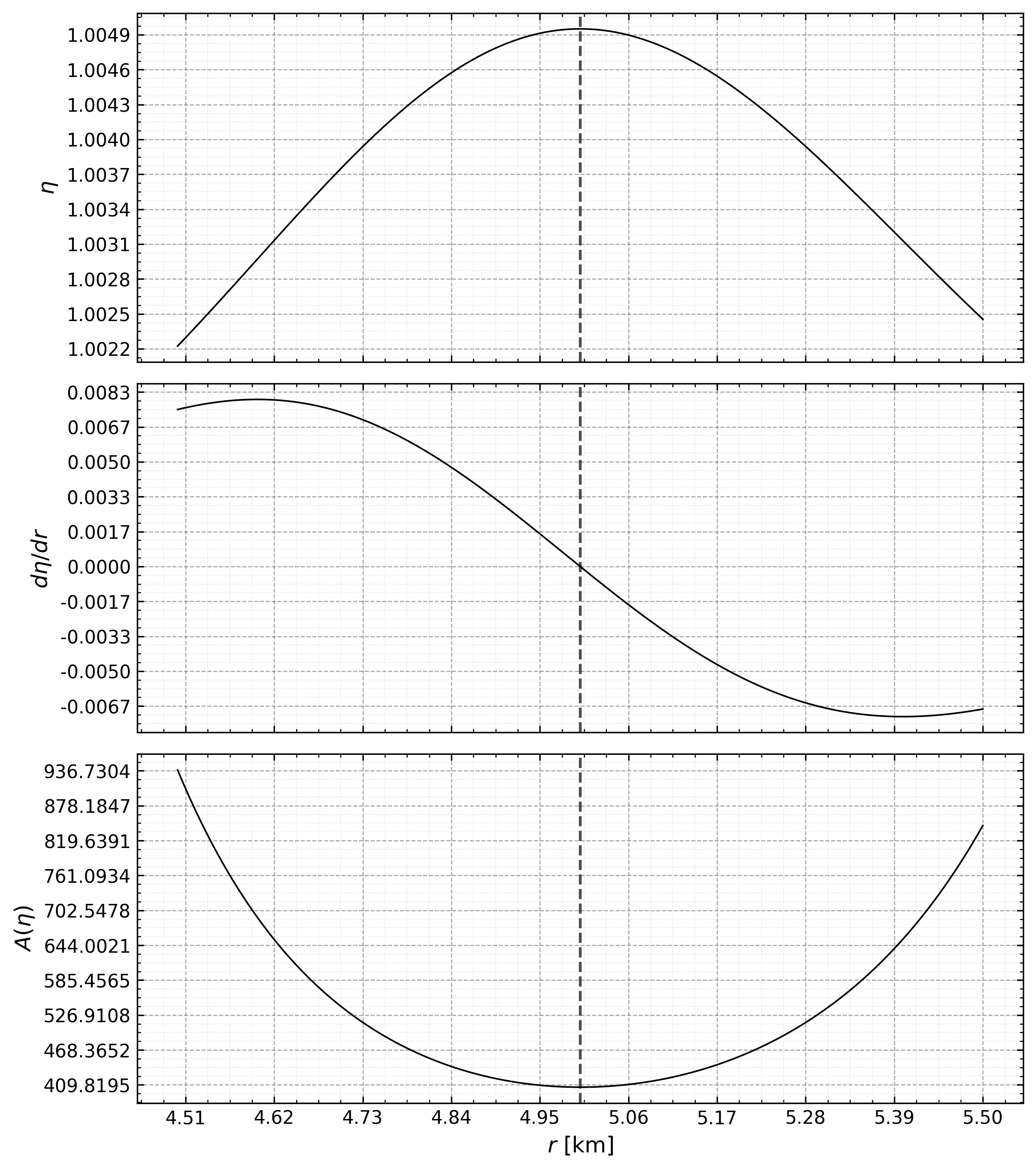}
   \caption{Behavior of the dimensionless energy density parameter $\eta$, its radial derivative, and the effective coefficient $A(\eta)$ near $r_1$, using parameter sets from Table~\ref{tab:tabparam3_4} for $M=2.05\,M_\odot$ (left) and $M=0.99\,M_\odot$ (right). For illustration, $r_1=5$ km is adopted without loss of generality. The plots show that the density profile remains continuous across the blurred region, while $A(\eta)$ encodes the coupling between core and crust EoS.}
    \label{fig:lane_emden_mod}
\end{figure}

Although classical, it is still interesting to consider the behavior of the conductivity function $A(\eta)$. Figure~\ref{fig:lane_emden_mod} illustrates typical solutions of $\eta$, $d\eta/dr$, and $A(\eta)$ as functions of $r$ near $r_1$, using parameters from Table~\ref{tab:tabparam3_4} for two representative masses, $M=2.05\,M_\odot$ and $M=0.99\,M_\odot$. The continuity of $\eta$ is manifest, while $A(\eta)$ highlights the non-trivial coupling structure across the interface. The physical interpretation of $A(\eta)$ as effective conductivity reveals that the core and crust communicate via a smooth rather than an abrupt transition.

A particularly sensitive diagnostic of phase-transition-like behavior is the baryonic chemical potential. In a true first-order transition, $\mu$ remains constant along a coexistence line; in our analytic model, no such plateau occurs, but pronounced changes in slope or weak non-monotonicity still indicate regions where the compressibility drops sharply. We therefore compute $\mu(\varepsilon)=(\varepsilon+p)/n_{B}$ for each regime. The resulting curves show that the most significant deviations from monotonicity occur in the parameter
regions where two positive critical densities exist, in agreement with the classification of Fig.~\ref{fig:mapa_parametros}. Interestingly, these signatures arise at densities well above the matching density $\varepsilon=\varepsilon_{0}$, meaning that the transition-like behavior originates deep in the core.

To further investigate possible phase transitions, we consider the baryonic chemical potential. For a cold NS, the first law of thermodynamics yields
\begin{equation}
    d\varepsilon = \mu\,dn_B,\qquad p = n_B\mu - \varepsilon,
\end{equation}
where $\mu$ is the baryon chemical potential, and $n_B$ is the baryon number density. Combining these,
\begin{equation}
    \frac{dn_B}{n_B} = \frac{d\varepsilon}{\varepsilon + p(\varepsilon)},
\end{equation}
whose integration gives $n_B(\varepsilon)$ up to a multiplicative constant. The chemical potential is then
\begin{equation}
    \mu(\varepsilon) = \frac{\varepsilon + p(\varepsilon)}{n_B(\varepsilon)}.
\end{equation}

In a first-order phase transition at $T=0$, one expects characteristic signatures such as plateaus or non-monotonic behavior in $\mu(\varepsilon)$: as energy is added, the system may remain at nearly constant $\mu$ while two phases coexist. Moreover, the derivative $d\mu/d\varepsilon$ may become small or even change sign in the coexistence region.

\begin{figure}[t]
    \centering
    \includegraphics[width=7.5cm,height=4.5cm]{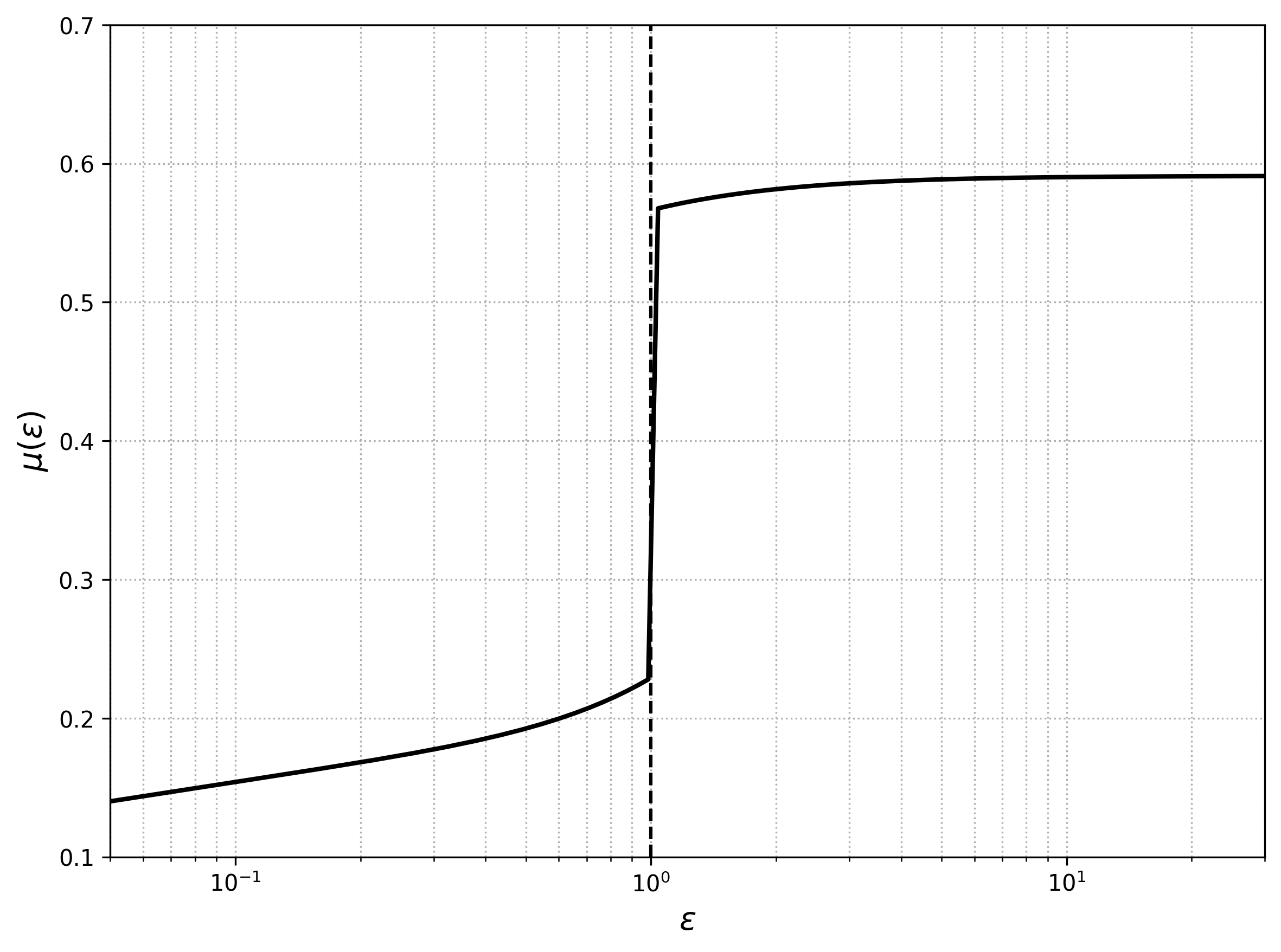}
    \includegraphics[width=7.5cm,height=4.5cm]{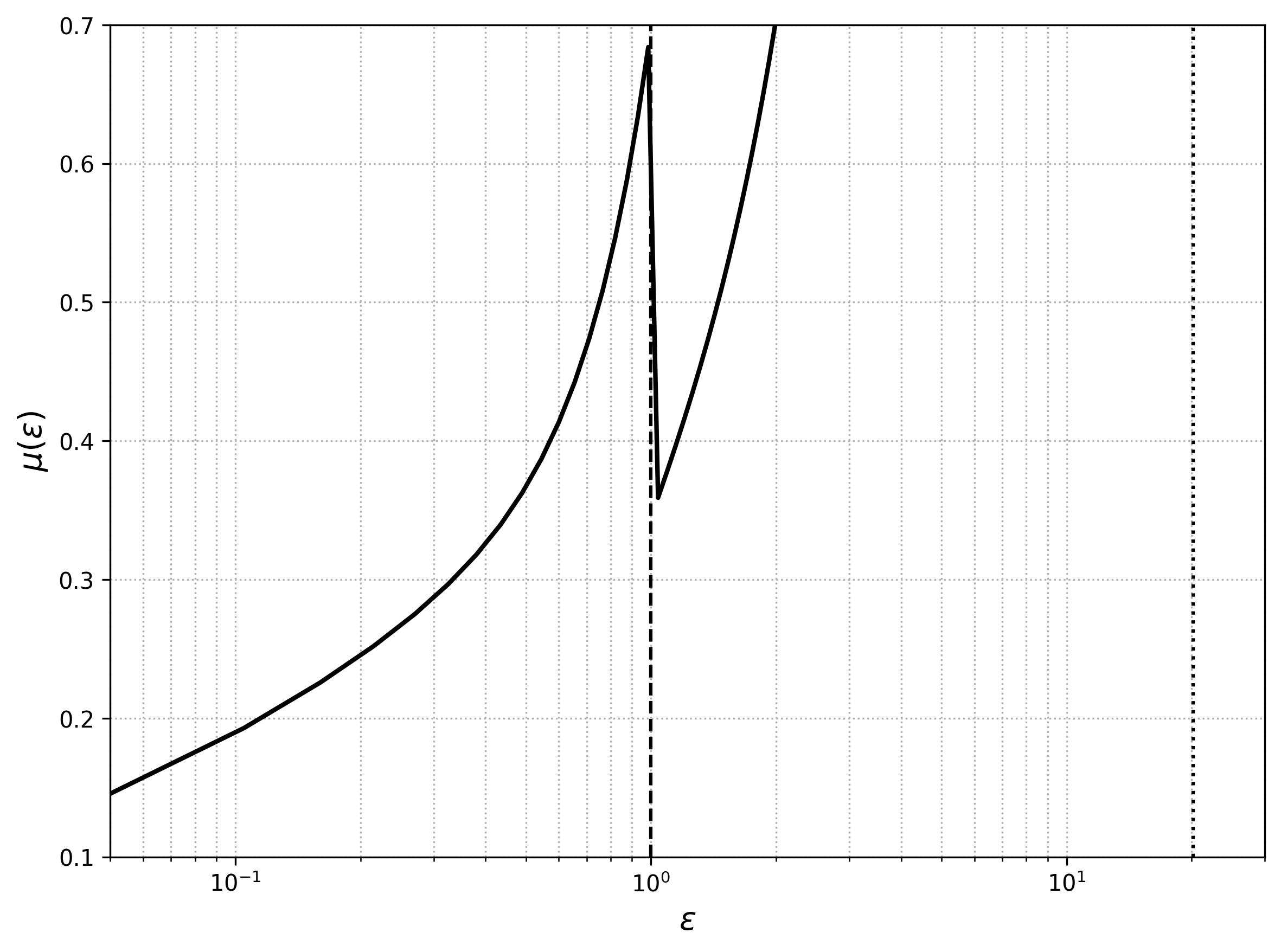}    
    \caption{Chemical potential $\mu(\varepsilon)$ for two representative NS configurations: $M=2.05\,M_\odot$ and $M=0.99\,M_\odot$, using parameter sets from Table~\ref{tab:tabparam3_4}. The most pronounced features in $\mu(\varepsilon)$ occur at densities significantly larger than the naive core–crust matching density $\varepsilon_0$, indicating that any effective phase transition is localized deep in the core rather than at the interface.}
    \label{fig:mu}
\end{figure}

\begin{figure}
    \centering
    \includegraphics[width=7.5cm,height=4.5cm]{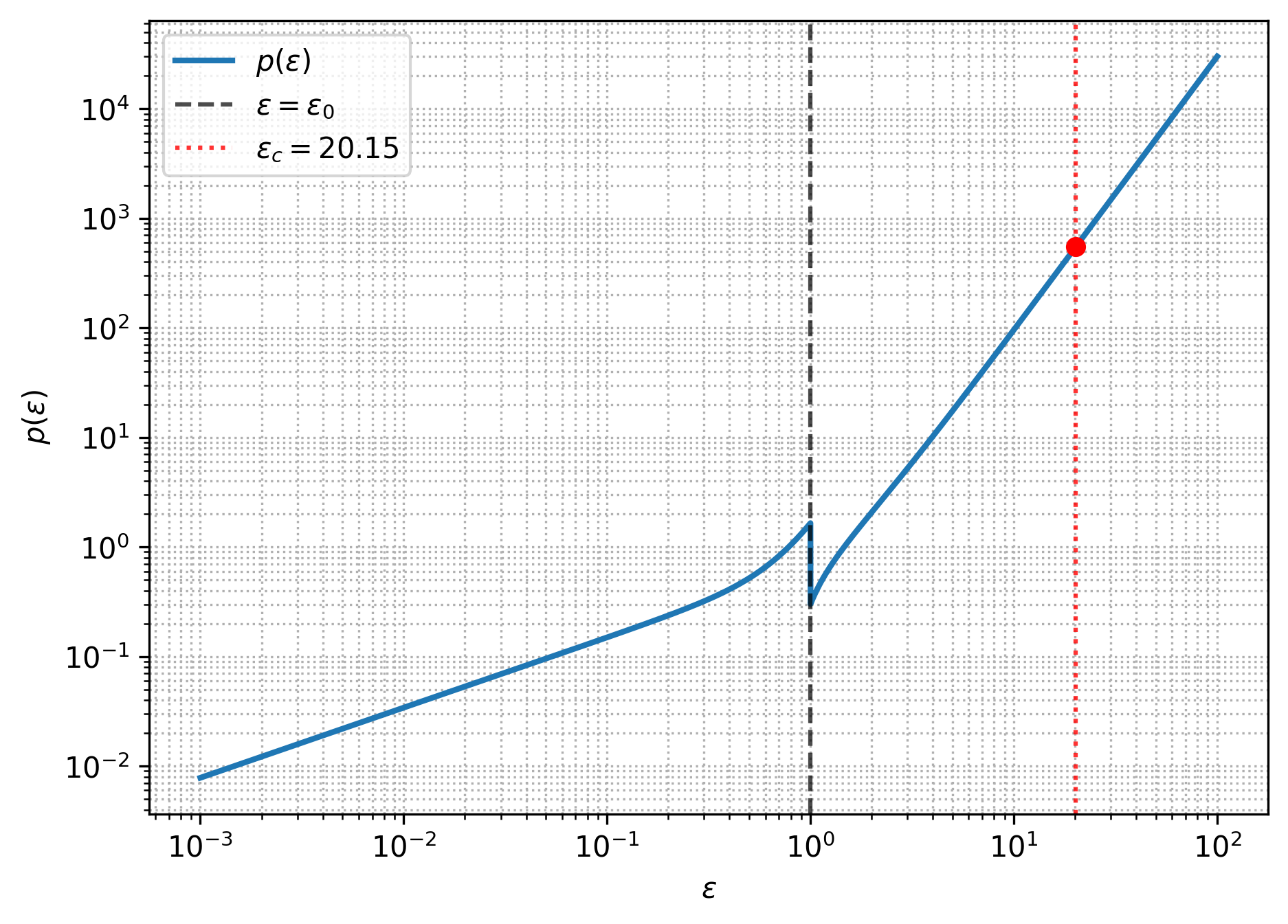}
    \includegraphics[width=7.5cm,height=4.5cm]{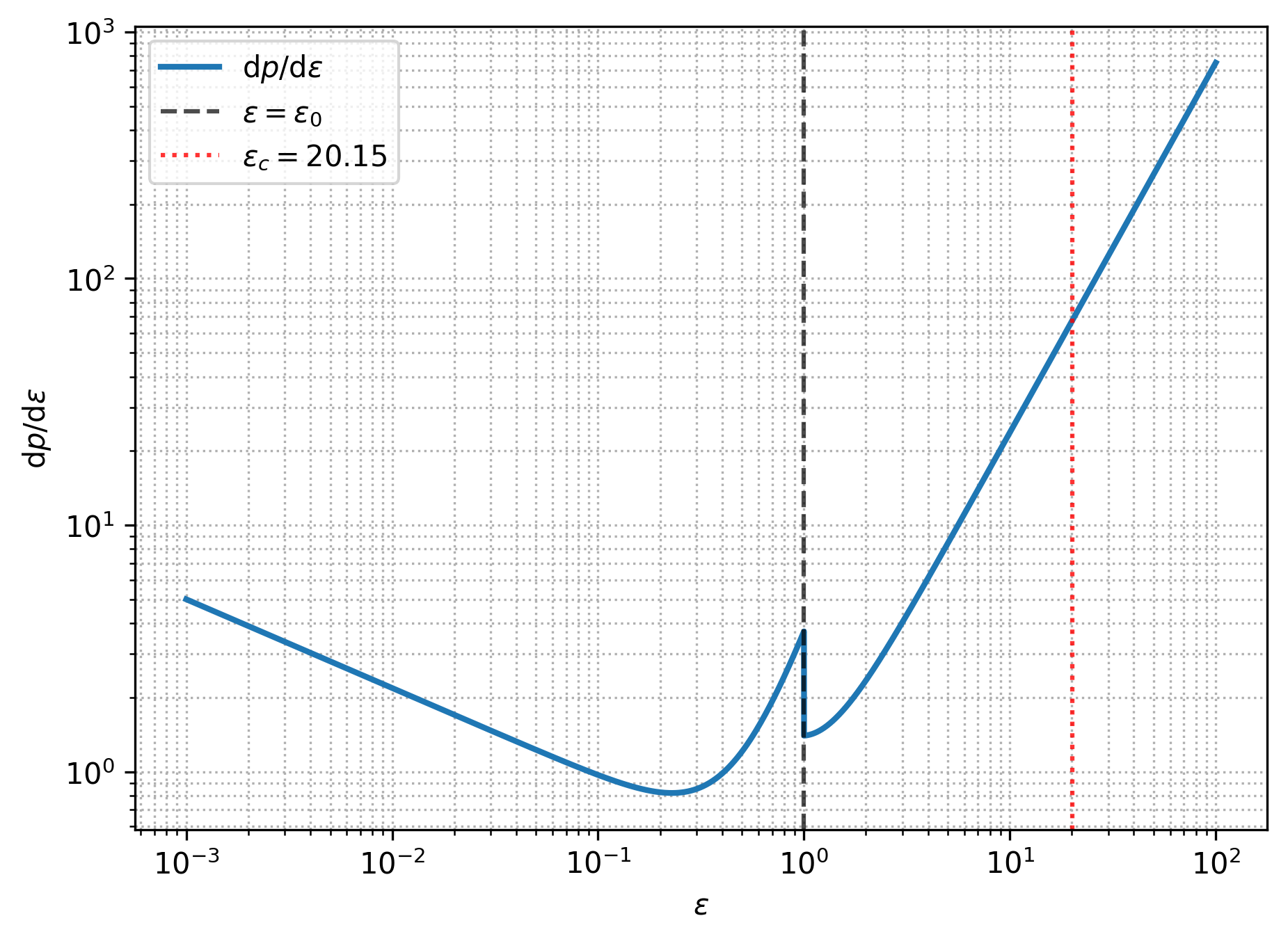}
    \includegraphics[width=7.5cm,height=4.5cm]{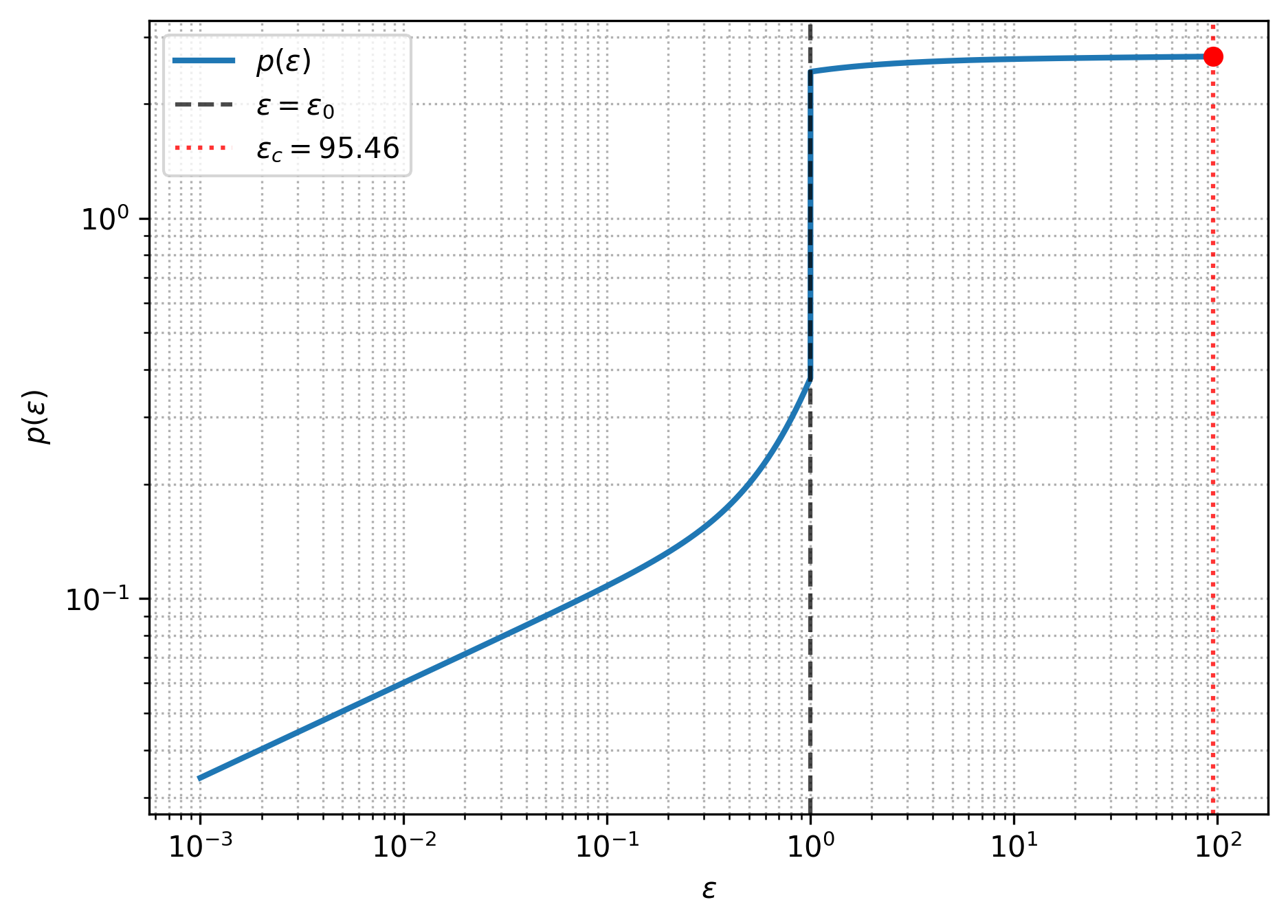}
    \includegraphics[width=7.5cm,height=4.5cm]{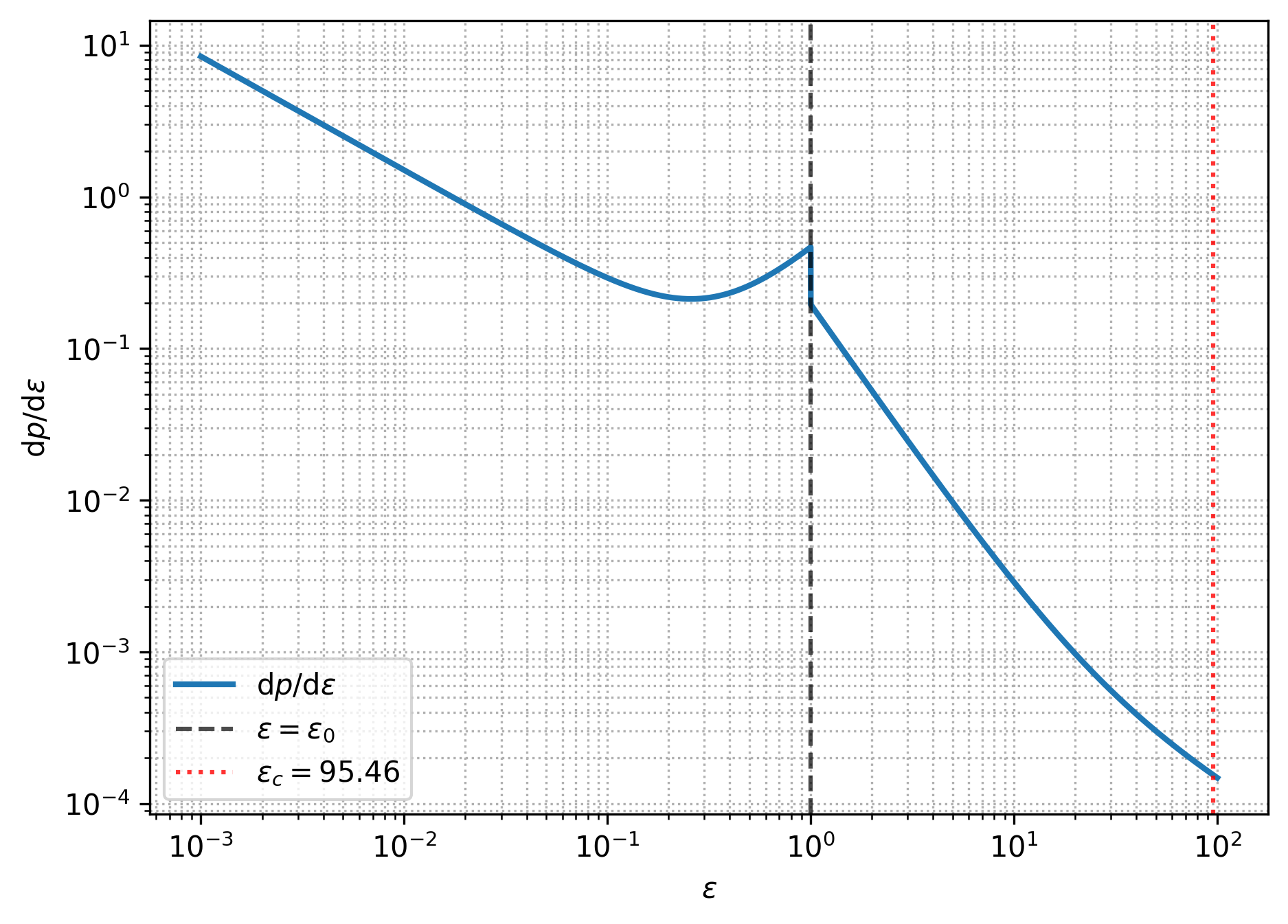}
    \caption{Upper panels represent the pressure and $dp/d\varepsilon$ for $M=2.05M_{\odot}$ while lower panels are for $M=0.99M_{\odot}$.} 
    \label{fig:pressure}
\end{figure}

Figure~\ref{fig:mu} presents the chemical potential as a function of $\varepsilon$ for two representative configurations in the regimes of Table~\ref{tab:tabparam3_4}, namely $M=2.05\,M_\odot$ and $M=0.99\,M_\odot$. The curves are computed using the piecewise EoS and the corresponding TOV solutions. Figure \ref{fig:pressure} shows the pressure and $dp/d\varepsilon$ as functions of $\varepsilon$. 

We find that $\mu(\varepsilon)$ is smooth and monotonically increasing across the entire density range, with no explicit plateaus or discontinuities. The strongest curvature changes occur at densities well above $\varepsilon_0$, i.e., deep in the core. This demonstrates that the thermodynamic signature of a potential phase transition does \emph{not} coincide with the artificial core–crust matching point defined by the piecewise EoS. Instead, the most significant nonlinear behavior arises at higher densities, where the vdW-like core term dominates. The absence of explicit plateaus in $\mu(\varepsilon)$ is consistent with Alford \textit{et al.} \cite{Alford.2013,Alford.2015}, showing that weak transitions can generate only curvature changes.

In the present formulation, the conductivity function $A(\eta)$ does more than ensure the continuity of the density profile across the blurred interface. Because it multiplies the highest-order derivative in Eq. \eqref{eq:aeta}, the sign and magnitude of  $A(\eta)$ determine whether the interface equation remains elliptic and, thus, whether the transition layer is mechanically stable. For all parameter sets in Tables \ref{tab:tabparam1}–\ref{tab:tabparam3_4}, we find  $A(\eta)>0$, so the problem is well posed and the core–crust interface behaves as a thin diffusive boundary layer rather than a discontinuity. Moreover, the characteristic width of this layer scales as $\ell^2\sim A(\eta)/(4\pi\eta)$. This scaling implies that larger values of  $A(\eta)$ lead to smoother, more extended transitions, whereas smaller values produce sharper gradients. In the regimes with two positive critical densities, $A(\eta)$ typically develops a pronounced minimum near $r_1$ (Fig. \ref{fig:lane_emden_mod}), reflecting a local reduction of the effective conductivity exactly where the competition between the vdW-like core term and the polytropic crust term is strongest. Thus, the radial structure of $A(\eta)$ provides a macroscopic tracer of the underlying phase-transition-like behavior, even though the strongest thermodynamic signatures in $\mu(\varepsilon)$ occur at densities well above the matching point.

Therefore, the piecewise structure of the EoS should not be directly identified with an actual thermodynamic phase boundary. The true ``transition'' region, as diagnosed by the behavior of the chemical potential, typically sits at much higher densities, possibly corresponding to changes in the microscopic composition (e.g., the onset of exotic degrees of freedom) or to strong softening/stiffening of the EoS.

The model's high sensitivity to parameter choices yields a wide range of stable configurations, including low-mass NSs. In particular, our parameter scans show that it is possible to obtain configurations with $M\simeq1.04\,M_\odot$ and $R\approx 11$ km, in line with recent low-mass NS candidates \cite{V.Doroshenko.NatureAstronomy.1-9.2022}. Nevertheless, the observed population of $0.7$–$1.0\,M_\odot$ neutron stars remains sparse, and both the formation channels (e.g., binary evolution) and the relevant microphysics are still uncertain. Therefore, any conclusions relying heavily on this mass range must be interpreted with caution.

\section{Discussion and Conclusions}
\label{sec:final}

Our results demonstrate that even highly simplified analytic equations of state can exhibit a surprisingly rich phenomenology. By explicitly mapping the core parameters $(\tau_{1},\sigma_{1})$, we showed that changes in the curvature of $p(\varepsilon)$ naturally divide the model into distinct physical regimes, ranging from monotonic behavior to domains with two effective critical densities. The generalized Lane–Emden formulation further revealed that the core–crust interface behaves as a thin diffusive layer whose properties depend sensitively on the curvature of the EoS. When embedded in the TOV equations, these features produce mass–radius relations that extend into the $0.7$–$1.1\, M_{\odot}$ range and generate thermodynamic signatures commonly associated with weak or smooth phase transitions.

Our analysis proceeded along two complementary lines. First, we examined the non-relativistic limit and derived generalized Lane–Emden–type equations for both the core and crust, including an effective description of the blurred core–crust transition. We showed that, near the transition radius $r_1$, the density profile satisfies an equation in which a density-dependent coefficient $A(\eta)$ acts as a generalized conductivity. This formulation emphasizes that the interface region is not a sharp discontinuity but a narrow layer where both branches of the EoS are effectively sampled. The resulting first-order system provides a convenient qualitative picture of how the density and its gradient behave across the interface.

Second, we solved the relativistic TOV equations using the piecewise EoS and systematically explored the parameter space $(\alpha_1,\beta_1,\tau_1,\sigma_1,\alpha_2,\beta_2,\tau_2,\sigma_2)$, constrained by thermodynamic stability, causality, and the existence of real, positive critical densities. We identified several regimes in the $(\tau_1,\sigma_1)$ plane, summarized in Table~\ref{tab:regimes}, and presented explicit parameter sets in Tables~\ref{tab:tabparam1}–\ref{tab:tabparam3_4} that produce realistic NS configurations with a variety of masses and radii. The corresponding mass–radius relations, shown in Figs.~\ref{fig:tov1}–\ref{fig:tov3_4}, illustrate how the interplay between thermal and interaction terms controls the stiffness of the EoS and, thus, the maximum mass.

An interesting outcome is that certain parameter regimes yield stable stars with gravitational masses around $M\simeq1.0$–$1.1\,M_\odot$. Such low-mass configurations are relevant in light of recent observational claims of very light NSs \cite{V.Doroshenko.NatureAstronomy.1-9.2022,J.E.Horvath.AA.672.L11.2023,J.Lin.2023}, as well as supernova simulations suggesting that remnants with $M\lesssim 1.2\,M_\odot$ can form under specific conditions \cite{B.Muller.Phys.Rev.Lett.134.071403.2025}. Although our EoS is intentionally simplified and not tuned to detailed microphysics, the ability to reproduce such low masses indicates that even phenomenological models with nonlinear density dependences can serve as proxies for more complex scenarios.

We have also computed the baryonic chemical potential as a function of the energy density for selected parameter sets, focusing on regimes where the EoS exhibits strong nonlinearities. The resulting $\mu(\varepsilon)$ curves, shown in Fig.~\ref{fig:mu}, do not display clear plateaus or discontinuities characteristic of textbook first-order phase transitions; instead, they show smooth, monotonic growth with pronounced curvature changes at densities significantly higher than the nominal core–crust matching density $\varepsilon_0$. This suggests that the ``physical'' transition region, if present, is driven by the nonlinear vdW-like core behavior and is located deep inside the star, rather than coinciding with the artificial piecewise junction of the EoS.

Our results emphasize that:

\begin{itemize}
    \item Piecewise EoS, even when built from relatively simple functional forms, can reproduce non-trivial thermodynamic behavior and generate rich families of mass–radius relations.
    \item The location of a phase-transition feature, as diagnosed by the chemical potential, does not necessarily match the EoS junction point; instead, it reflects the internal competition between different pressure contributions at high density.
    \item Low-mass NS configurations ($M\lesssim1.1\,M_\odot$) are compatible with our simplified model and thus cannot be ruled out solely based on gross structural arguments. Their actual existence in nature, however, depends on the dynamics of supernovae and the microphysics of dense matter.
\end{itemize}

Several extensions of this work are natural. A more realistic treatment could include finite-temperature effects, rotation, magnetic fields, and multi-component compositions (hyperons, quarks, meson condensates). One could also confront the model with current multi-messenger constraints from gravitational-wave events and X-ray timing (e.g., NICER), which increasingly restrict allowable EoS. Finally, embedding the present phenomenological EoS into more detailed microphysical frameworks could help clarify which features of our piecewise construction are robust and which are artifacts of the simplified parametrization.

Although the model is not microphysical, it provides a controlled setting in which the impact of nonlinearities in $p(\varepsilon)$ can be systematically traced from the equation of state to the stellar structure and the chemical potential. This makes it a useful testbed for studying low-mass neutron stars and for exploring how effective phase-transition features arise in analytic or semi-analytic EoS models.

\section*{Acknowledgments}

PHFA and SDC acknowledge the Federal University of São Carlos (UFSCar) and the Applied Mathematics Laboratory (CCTS/DFQM) for their institutional support and stimulating research environment.




\begin{thebibliography}{99}

\bibitem{R.C.Tolman.1934}R. C. Tolman, \emph{Effect of inhomogeneity on cosmological models}, \emph{Proc. Nat. Acad. Sci.} {\bf 20} (1934) pg. 3

\bibitem{R.C.Tolman.1939}R. C. Tolman, \emph{Static solutions of Einstein’s field equations for spheres of fluid}, \emph{Phys. Rev.} {\bf 55} (1939) pg. 364

\bibitem{R.J.Oppenheimer.G.M.Volkoff.1939}J. R. Oppenheimer and G. M. Volkoff, \emph{On massive neutron cores}, \emph{Phys. Rev.} {\bf 55} (1939) pg. 334

\bibitem{E.Fermi.1926}E. Fermi, \emph{Zur Quantelung des idealen einatomigen Gases}, \emph{Zeitschrift f\"ur Physik} {\bf 36} (1926) pg. 902

\bibitem{C.G.Bassa.et.al.2017}C. G. Bassa {\it{et al.}}, \emph{LOFAR discovery of the fastest-spinning millisecond pulsar in the galactic field}, \emph{Astrophys. J. Lett.} {\bf 846} (2017) pg. L20

\bibitem{J.Antoniadis.et.al.2013}J. Antoniadis {\it{et al.}}, \emph{A massive pulsar in a compact relativistic binary}, \emph{Science} {\bf 340} (2013) pg. 1233232

\bibitem{M.Linares.Astrophys.J.2018}M. Linares, T. Shahbaz and J. Casares, \emph{Peering into the dark side: magnesium lines establish a massive neutron star in PSR J2215+5135}, \emph{Astrophys. J.} {\bf 859} (2018) pg. 1

\bibitem{Miller.2019}M. C. Miller \textit{et al.}, \emph{PSR J0030+0451 mass and radius from NICER data and implications for the properties of neutron star matter}, \emph{Astrophys. J. Lett.} {\bf 887} (2019) pg. L24

\bibitem{Riley.2019}T. E. Riley \text{et al.}, \emph{A NICER view of PSR J0030+0451: millisecond pulsar parameter estimation}, \emph{Astrophys. J. Lett.} {\bf 887} (2019) pg. L21

\bibitem{Miller.2021}M. C. Miller \textit{et al.}, \emph{The radius of PSR J0740+6620 from NICER and XMM-Newton data}, \emph{Astrophys. J. Lett.} {\bf 918} (2021) pg. L28

\bibitem{Riley.2021}T. E. Riley \textit{et al.}, \emph{A NICER view of the massive pulsar PSR J0740+6620 informed by radio timing and XMM-Newton spectroscopy}, \emph{Astrophys. J. Lett.} {\bf 918} (2021) pg. L27

\bibitem{V.Doroshenko.NatureAstronomy.1-9.2022}V. Doroshenko \textit{et al.}, \emph{A strangely light neutron star within a supernova remnant}, \emph{Nature Astronomy} {\bf 6} (2022) pg. 1444

\bibitem{J.E.Horvath.AA.672.L11.2023}J. E. Horvath \textit{et al.}, \emph{A light strange star in the remnant HESS J1731-347: Minimal
consistency checks}, \emph{A$\&$A} {\bf 672} (2023) pg. L11

\bibitem{J.Lin.2023}J. Lin \textit{et al.}, \emph{An X-Ray-dim ``isolated'' neutron star in a binary?}, \emph{Astrophys. J. Lett.} {\bf 944} (2023) pg. L4

\bibitem{B.Muller.Phys.Rev.Lett.134.071403.2025}B. Müller, A. Heger and J. Powell, \emph{Minimum neutron star mass in neutrino-driven supernova explosions}, \emph{Phys. Rev. Lett.} {\bf 134} (2025) pg. 071403

\bibitem{Wang.2024}T. Wang and A. Burrows, \emph{Supernova explosions of the lowest-mass massive star progenitors}, \emph{Astrophys. J.} {\bf 969} (2024) pg. 74

\bibitem{Abbott.2018}B. P. Abbott \textit{et al.}, \emph{GW170817: measurements of neutron star radii and equation of state}, \emph{Phys. Rev. Lett.} {\bf 121} (2018) pg. 161101

\bibitem{Abbott.2019}B. P. Abbott \textit{et al.}, \emph{Binary black hole population properties inferred from the first and second observing runs of advanced LIGO and advanced Virgo}, \emph{Astrophys. J. Lett.} {\bf 882} (2019) pg. L24

\bibitem{A.G.W.Cameron.1959}A. G. W. Cameron, \emph{Neutron star models}, \emph{Astrophys. J.} {\bf 130} (1959) pg. 884

\bibitem{G.Baym.1969}G. Baym, C. Pethick, and D. Pines, \emph{Superfluidity in neutron stars}, \emph{Nature} {\bf 224} (1969) pg. 673

\bibitem{G.Baym.1971}G. Baym, C. Pethick, and P. Sutherland, \emph{The ground state of matter at high densities: equation of state and stellar models}, \emph{Astrophys. J.} {\bf 170} (1971) pg. 299

\bibitem{A.Akmal.1998}A. Akmal, V. R. Pandharipande, and D. G. Ravenhall, \emph{Equation of state of nucleon matter and neutron star structure}, \emph{Phys. Rev. C} {\bf 58} (1998) pg. 1804

\bibitem{A.R.Raduta.2021}A. R. Raduta, F. Nacu, and M. Oertel, \emph{EoS for hot neutron stars}, \emph{Eur. Phys. J. A} {\bf 57} (2021) pg. 329

\bibitem{Z.Ji.2025}Z. Ji and J. Chen, \emph{The equation of state of neutron stars: theoretical models, observational constraints, and future perspectives}, arXiv:2502.05513v1

\bibitem{J.Baacke.1977}J. Baacke, \emph{Thermodynamics of a gas of MIT bags}, \emph{Acta Phys. Pol.} {\bf 88} (1977) pg. 625

\bibitem{A.Chodos.1974}A. Chodos, R. L. Jaffe, K. Johnson, C. B. Thorn and V. F. Weisskopf, \emph{A new extended model of hadrons}, \emph{Phys. Rev. D} {\bf 9} (1974) pg. 3471

\bibitem{Typel.2016}D. Alvarez-Castillo, S. Benic, D. Blaschke, S. Han and S. Type, \emph{Neutron star mass limit at 2 $M_{\odot}$ supports the existence of a CEP}, \emph{Eur. Phys. J. A} {\bf 52} (2016) pg. 232

\bibitem{V.Vovchenko.2020}V. Vovchenko, \emph{Hadron resonance gas with van der Waals interactions}, \emph{Int. J. Mod. Phys. E} {\bf 29} (2020) pg. 2040002

\bibitem{verma.arxiv.2025}A. Verma, A. K. Saha and R. Mallick, \emph{Comparison of equations of state for neutron stars with first-order phase transitions: a qualitative study}, \emph{Astrophys. J.} {\bf 985} (2025) pg. 1

\bibitem{Alford.2013}M. G. Alford, S. Han and M. Prakash, \emph{Generic conditions for stable hybrid stars}, \emph{Phys. Rev. D} {\bf 88} (2013) pg. 083013

\bibitem{Alford.2015}M. G. Alford, G. F. Burgio, S. Han, G. Taranto and D. Zappalà, \emph{Constraining and applying a generic high-density equation of state}, \emph{Phys. Rev. D} {\bf 92} (2015) pg. 083002

\end{thebibliography}

\end{document}